\documentclass[prd,aps,eqsecnum,amsmath,floatfix,nofootinbib,preprint,tightenlines]{revtex4}

\usepackage{latexsym}
\usepackage{graphicx}
\usepackage[dvipsnames]{xcolor}
\usepackage{amssymb}

\def\bibi{\bibitem}
\def\floatcaption#1#2{ \caption{#2 \label{#1}} }

\let\inodot=\i

\def\a{\alpha}

\def\c{\chi}

\def\g{\gamma}

\def\i{\iota}

\def\s{\sigma}                  
\def\t{\tau}

\def\D{\Delta}

\def\P{\Pi}

\def\cbo{{\,\raise-.15ex\Sc [\,}}                       
\def\gtap{\raisebox{-.4ex}{\rlap{$\sim$}} \raisebox{.4ex}{$>$}}   

\def\ddt#1{{\buildrel {\hbox{\LARGE .\kern-2pt.}} \over {#1}}}

\def\ie{\mbox{\it i.e.}}

\long\def\symbolfootnote[#1]#2{\begingroup%
\def\thefootnote{\fnsymbol{footnote}}\footnote[#1]{#2}\endgroup}

\long \def \blockcomment #1\endcomment{}

\def\seef{{\it cf.}}

\begin{document}

\thispagestyle{empty}

\begin{center}
\vspace*{5mm}
\begin{boldmath}

\vspace{.5cm}
{\large\bf Evidence against naive truncations of the OPE from $e^+e^- \to$~hadrons below charm}
\end{boldmath}
\\[10mm]
Diogo Boito,$^a$
Maarten Golterman,$^{b}$
Kim Maltman,$^{c,d}$
Santiago Peris$^e$
\\[8mm]
{\small\it
$^a$Instituto de F{\'\inodot}sica de S{\~a}o Carlos, Universidade de S{\~a}o Paulo\\
CP 369, 13570-970, S{\~a}o Carlos, SP, Brazil
\\[5mm]
$^b$Department of Physics and Astronomy,
San Francisco State University\\ San Francisco, CA 94132, USA
\\[5mm]
$^c$Department of Mathematics and Statistics,
York University\\  Toronto, ON Canada M3J~1P3
\\[5mm]
$^d$CSSM, University of Adelaide, Adelaide, SA~5005 Australia
\\[5mm]
$^e$Department of Physics and IFAE-BIST, Universitat Aut\`onoma de Barcelona\\
E-08193 Bellaterra, Barcelona, Spain
}
\\[10mm]
\end{center}

\begin{quotation}
The operator product expansion (OPE), truncated in dimension, is
employed in many contexts.   An example is the extraction of the
strong coupling, $\a_s$, from hadronic $\t$-decay data, using a variety of
analysis methods based on finite-energy sum rules.
Here, we reconsider a long-used method,
which
parametrizes non-perturbative contributions to the $I=1$ vector and axial
vacuum polarizations with the OPE, setting
several higher-dimension coefficients
to zero in order to implement the method in practice.
The assumption that doing this has a negligible effect on the value of
$\a_s$ is tantamount to the assumption that the low-dimension part
of the OPE converges rapidly with increasing
dimension near the $\t$ mass. Were this assumption
valid, it would certainly have to be valid at energies above the $\t$ mass
as well. It follows that the method can be tested using data
obtained from $e^+e^-\to\mbox{hadrons}$, as they are not limited by the
kinematic constraints of $\t$ decays. We carry out such an investigation
using a recent high-precision compilation for the $R$-ratio, arguing
that it provides insights into the validity of the strategy,
even if it probes a different, though related channel. We find
that $e^+e^-$-based tests call into question the implied assumption of
rapid convergence of the low-dimension part of the OPE
around the $\tau$ mass, and thus underscore the need to
restrict finite-energy sum-rule analyses to observables which receive only contributions
from lower-order terms in the OPE.
\end{quotation}

\newpage
\section{\label{introduction} Introduction}
As is well known, the spectral function, $\rho_{\rm EM}(s)$, of
$\Pi_{\rm EM}$, the scalar polarization of the electromagnetic (EM)
current-current two-point function, is directly obtainable from the
experimentally measured $R$-ratio,
\begin{eqnarray}
&&R(s)\equiv{\frac{3s}{4\pi\alpha^2}}\,\sigma_{e^+e^-\to{\rm hadrons}
(\gamma)}(s) ={\frac{\sigma_{e^+e^-\to{\rm hadrons}(\gamma)}(s)}
{\sigma_{e^+e^-\to\mu^+\mu^-}(s)}}\ ,
\label{Rdefn}\end{eqnarray}
via
\begin{eqnarray}
&&\rho_{\rm EM}(s)={\frac{1}{\pi}}\,\mbox{Im}\,\Pi_{\rm EM}(s)=
{\frac{1}{12\pi^2}}\,R(s)\ ,
\label{rhoemfromR}\end{eqnarray}
where, in Eq.~(\ref{Rdefn}), $\a$ is the fine-structure constant, the
second of the equations holds for values of $s$ for which the muon mass
can be neglected, and the $\gamma$ in parentheses indicates that
the hadronic states in question are inclusive of final-state radiation.

Similarly, information on the spectral functions, $\rho^{(J)}_{V/A;ij}(s)$,
of the spin $J=0,\, 1$ scalar polarizations, $\Pi_{ij;V/A}^{(J)}$, of the
flavor $ij=ud$ and $us$ vector (V) and axial vector (A) current-current
two-point functions can be obtained from the experimental hadronic $\tau$-decay
distributions, $dR_{V/A;ij}/ds$. Explicitly~\cite{tsai71},
\begin{eqnarray}
\label{diffR}
{\frac{dR_{V/A;ij}(s;s_0)}{ds}}&=&
{\frac{12\pi^2\vert V_{ij}\vert^2 S_{\rm EW}}
{s_0}}\, \left[ w_\tau (y_\tau ) \rho_{V/A;ij}^{(0+1)}(s)
- w_L (y_\tau )\rho_{V/A;ij}^{(0)}(s) \right]\ ,
\label{basictaudecay}
\end{eqnarray}
where $y_\tau =s/m_\tau^2$,
\begin{eqnarray}
\label{wtau}
w_\tau (y)&=&(1-y)^2(1+2y)\ ,\\
w_L(y)&=&2y(1-y)^2\ ,\nonumber
\end{eqnarray}
$S_{\rm EW}$ is a known short-distance electroweak correction~\cite{erler},
$V_{ij}$ is the flavor $ij$ CKM matrix element{\footnote
{Eq.~(\ref{basictaudecay}) has been written in terms of spectral function
combinations, $\rho_{V/A;ij}^{(0+1)}(s)$ and $s\rho_{V/A;ij}^{(0)}$, for
which the corresponding polarizations, $\Pi_{V/A;ij}^{(0+1)}(s)$ and
$s\, \Pi_{V/A;ij}^{(0)}(s)$, are free of kinematic singularities.}}
and $dR_{V/A;ij}(s;s_0)/ds$ is related to the total inclusive
hadronic $\tau$-decay width by
\begin{eqnarray}
\label{tauRratiodefn}
R_{V/A;ij}(s_0)&=&\int_0^{s_0}ds\,{\frac{dR_{V/A;ij}(s;s_0)}{ds}}\ , \\
R_{V/A;ij}(m_\t^2)&=&\frac{\Gamma
[\tau^- \rightarrow \nu_\tau \, {\rm hadrons}_{V/A;ij}\, (\gamma )]}{
\Gamma [\tau^- \rightarrow \nu_\tau e^- {\bar \nu}_e (\gamma)]}\ .\nonumber
\end{eqnarray}

The analyticity properties of current-current polarizations (denoted
generically by $\Pi$) ensure the validity of finite-energy sum rules
(FESRs), which allow one to relate weighted integrals over the associated
experimental spectral
data to theoretical representations of the polarizations~\cite{fesrrefs}.
Explicitly, for any $\Pi (s)$ free of kinematic singularities, any
$s_0>0$, and any $w(s)$ analytic inside and on the contour
$\vert s\vert =s_0$, one has the sum rule
\begin{equation}
\label{basicfesr}
I_w^{\rm exp}(s_0)=I_w^{\rm th}(s_0)\ ,
\end{equation}
where the weighted integrals over the experimental spectral function and
over the vacuum polarization are defined as
\begin{subequations}
\begin{eqnarray}
\label{basicfesrdefs}
I_w^{\rm exp}(s_0)&=&\frac{1}{s_0}\int_0^{s_0}\, ds\,
w\left(\frac{s}{s_0}\right)\, \rho (s)\ ,
\label{basicfesrdefsa}\\
I_w^{\rm th}(s_0) &=& -{\frac{1}{2\pi i s_0}}\, \oint_{\vert s\vert =s_0}
ds\,w\left(\frac{s}{s_0}\right)\, \Pi (s)\ .\label{basicfesrdefsb}
\end{eqnarray}
\end{subequations}

For sufficiently large $s_0$, $I_w^{\rm th}(s_0)$ can be
approximated using the operator product expansion (OPE) for $\Pi$. This
allows quantities entering the OPE (such as the strong coupling
$\alpha_s$, quark masses,
and effective higher-dimension vacuum condensates) to be related to
experimental data, in principle. FESRs based on $ud-us$ flavor-breaking
differences of hadronic $\tau$-decay distributions can also be used to
provide an independent determination of
$\vert V_{us}\vert$~\cite{gamizfbfesrvus,kmcwfbfesrvus,hlmz17}.
Generalizing Eq.~(\ref{basicfesr}) to weights $w(s)/s^N$, still with $w(s)$
analytic, yields analogous inverse-moment FESR (IMFESR) relations involving
quantities such as $\Pi (0)$, and its derivatives with respect to $s$
at $s=0$, which can be exploited to determine some of the low-energy
constants of chiral peturbation theory (ChPT), provided those terms
in the associated OPE required for the $w(s)$ chosen are known from
external sources~\cite{IMFESR}. The OPE thus plays an important role
in FESR and IMFESR analyses.

Information on the flavor $ud$, $us$ $V$ and $A$ spectral functions from
hadronic $\tau$-decay data is, of course, only available up to the
kinematic limit, $s=m_\tau^2$, restricting FESRs and IMFESRs based on
hadronic $\tau$-decay data to $s_0\le m_\tau^2$. No such kinematic limit
exists for FESRs and IMFESRs based on hadronic electroproduction cross-section
data.

The OPE is expected to provide an accurate representation of $\Pi(s)$ valid
for Euclidean $Q^2\, \equiv\, -s \gg\Lambda_{\rm QCD}^2$,
up to small exponentially suppressed corrections.
In Eq.~(\ref{basicfesrdefsb}), the OPE representation, however,
must be used over the whole of the contour $\vert s\vert =s_0$,
which includes the region near the Minkowski axis, where,
as anticipated in Ref.~\cite{pqw}, the OPE breaks down at intermediate
(timelike) $s$.  This is clear from the presence of resonance peaks
in experimental spectral functions at $s$ of order a few GeV$^2$.
Such ``duality violating'' (DV) effects are
expected to be localized to the vicinity of the Minkowski axis, an
expectation confirmed by studies of FESRs employing both ``unpinched''
weights (those which do not vanish at $s=s_0$ and hence do not suppress
contributions from the region near the Minkowski axis) and ``pinched''
weights (those with $w(s_0)=0$, which do suppress contributions from
that region)~\cite{kmpinching}. Precision determinations of
$\a_s$, quark masses, and other OPE parameters
may, however, require small residual DV contributions to be taken
into account, even for FESRs involving pinched weights \cite{CGP,alphas1}.

A second, related issue for the use of the OPE in FESRs and IMFESRs is
the fact that the OPE (an expansion in $z=1/Q^2$) is not
convergent \cite{BCGMP}. Convergence would require the existence of 
a region in the complex plane around $z=0$ free
of singularities, and hence, in the case of a current-current two-point
function, the vanishing of the corresponding spectral function
above some maximum value of $s$. This is not the case. The OPE is thus,
at best, an asymptotic expansion, and one cannot safely
assume that effective condensates $C_D$, of dimension $D$,
defined by\footnote{The
condensates are logarithmically dependent on $Q^2$. This dependence,
which is suppressed by at least one power of $\a_s$, is
usually neglected for $D\ge 4$, as it makes no difference in the
value of $\a_s$ obtained from FESR analyses \cite{alphas1}.}
\begin{equation}
\label{OPEdef}
\P_{\rm OPE}(-Q^2)=\sum_{k=0}^\infty\frac{C_{2k}(Q^2)}{(-Q^2)^k}\ ,
\end{equation}
naively scale as $\Lambda_{\rm QCD}^D$.

This is relevant for FESRs and IMFESRs employing weights $w(s)$
which generate OPE contributions proportional to higher dimension $C_D$
not known from external sources.  Assumptions based on naive scaling
of the $C_D$ have often been used to argue that such unknown contributions
are ``safely'' negligible at scales of a few GeV$^2$, including
$s_0\sim m_\tau^2$. In general, a polynomial weight
$w(y)=\sum_{k=0}^N b_k y^k$, $y=s/s_0$, produces, up to logarithmic corrections
suppressed by additional powers of $\a_s$, a contribution
\begin{eqnarray}
\sum_k\,(-1)^k b_k\, C_{2k+2}/s_0^{k+1}
\label{higherDfesrcont}
\end{eqnarray}
to the right-hand side of Eq.~(\ref{basicfesr}).

We refer to the prescription of neglecting contributions proportional
to $b_k\, C_{2k+2}$ for higher $k$ and $s_0$ of order a few GeV$^2$
as the ``truncated OPE'' (tOPE) approach \cite{BGMP16}. The non-convergence
of the OPE implies that this assumption is a dangerous one to
make, in general. It does, however, remain a logical possibility that,
for a given value of $s_0$, the truncated OPE might represent a
reasonable approximation for a specific set of weights.  If so, the
tOPE approach can be used for a sum-rule analysis employing this set of
weights at this value of $s_0$.
In this paper, we will investigate whether or not this is the case for the EM
current-current two-point function for values of $s_0$ between
$m_\tau^2$~and~$4$~GeV$^2$.

An example of a situation in which the tOPE approximation
might be practically useful is provided by the FESR determination
of $\alpha_s$ based on non-strange hadronic $\tau$ decay data.
Since the kinematic weight $w_\tau (y)=1-3y^2+2y^3$ appearing in
Eq.~(\ref{basictaudecay}) has degree $3$, the OPE representation
of the total non-strange hadronic $\tau$-decay width contains
contributions of dimension $0,\, 6$ and $8$.\footnote{$D=4$
contributions are strongly suppressed by the absence of a term
linear in $y$ in $w_\tau (y)$.  For the case of non-strange $\t$ decays,
$C_2$ is proportional
to the square of the light quark mass, and is numerically negligible.
For the case of the $R$-ratio, there is a contribution proportional to the 
square of the strange quark
mass which can be calculated; for details
we refer to Ref.~\cite{alphasEM}.} The total non-strange width
(corresponding to the kinematically weighted spectral integral
with $s_0=m_\tau^2$) is thus insufficient, by itself, to allow
one to determine $\alpha_s$ since the relevant condensates, $C_6$
and $C_8$ are not known from external sources. A tOPE strategy
to deal with this problem, proposed in Ref.~\cite{DibPich},
is to consider additional FESRs in which $C_6$ and $C_8$ also occur.
The conventional version of this strategy employs the five
``$k\ell$ spectral weights,''
\begin{equation}\label{spectralweights}
w_{k\ell}(y)=(1+2y)(1-y)^{2+k}y^l \ ,
\end{equation}
with $k\ell=00,10,11,12$ and $13$ (note that $w_{00}=w_\t$), and focuses on the
$s_0=m_\tau^2$ versions of the corresponding spectral integrals~(\ref{basicfesrdefsa}).
Recent versions of this analysis may be found in Refs.~\cite{ALEPH13,Pich}.
A number of alternate weight sets, including the so-called ``optimal
weights,''
\begin{equation}\label{optimalweights}
w_{2k}(y)=1-(k+2)y^{k+1}+(k+1)y^{k+2}\ ,
\end{equation}
$k=1,\cdots ,5$,
were also considered in Ref.~\cite{Pich} (note that $w_{21}=w_\t$),{\footnote{The absence of a term linear in $y$ again
strongly suppresses $D=4$ OPE contributions for the optimal weight
FESRs.}} with $s_0$ again restricted to
$m_\tau^2$.{\footnote{In Ref.~\cite{Pich}, $s_0$ dependence was considered, but
all final values quoted for $\a_s(m_\t^2)$ were obtained from moments at $s_0=m_\t^2$.}

The tOPE assumption enters these analyses as follows. Since both
the $k\ell$ spectral and optimal weight sets
involve weights with degrees up to $7$, OPE contributions
up to $D=16$ are in principle required, as per Eq.~(\ref{higherDfesrcont}). So long as one attempts
to minimize residual DV contributions by restricting $s_0$ to its
maximum kinematically allowed value, $m_\tau^2$, a five-weight set
provides only five (highly correlated) spectral integrals for use in fitting, and one
can hence fit at most four OPE parameters. The five $k\ell$ spectral weight
FESRs, however, in general, involve OPE contributions depending on
$\alpha_s$, and the seven condensates $C_4,\cdots ,C_{16}$.
The five optimal-weight FESRs, similarly, neglecting the strongly
suppressed $D=4$ contributions, depend on the OPE parameters $\alpha_s$
and $C_6,\cdots , C_{16}$. In both cases, the number of OPE parameters
exceeds the number of $s_0=m_\tau^2$ spectral integrals,
unless one makes the tOPE assumption, which is to neglect contributions
from the new $C_D$ introduced by the higher degree weights.
In the tOPE implementation of the conventional $k\ell$ spectral-weight
analysis, contributions proportional to $C_{10}$, $C_{12}$, $C_{14}$
and $C_{16}$ are assumed negligible, leaving the four remaining
OPE parameters $\alpha_s$, $C_4$, $C_6$ and $C_8$ to be fit.
In the tOPE implementation of the optimal-weight analysis,
contributions proportional to $C_{12}$, $C_{14}$ and $C_{16}$ are
assumed negligible and the five spectral integrals are used to fit
the four remaining relevant OPE parameters, $\alpha_s$, $C_6$, $C_8$
and $C_{10}$. In both cases, the assumption underlying the tOPE
approach is that the OPE, though not actually convergent, nonetheless
behaves, for $s_0=m_\tau^2$, as if it were a rapidly converging series
out to at least $D=16$.

Since both integrated DV contributions and integrated higher-dimension OPE
contributions decrease with increasing $s_0$, it follows that, if
the tOPE assumption is reliable at $s_0=m_\tau^2$, it should be even more
reliable for $s_0>m_\tau^2$. Unfortunately, the kinematic restriction
$s_0\le m_\tau^2$ prevents the self-consistency tests this suggests
from being carried out for $\tau$-based FESRs. Analogous tests can,
however, be carried out for FESRs based on EM $R$-ratio data, where
there is no kinematic restriction on the hadronic invariant mass-squared
$s$.\footnote{For other tests of the tOPE strategy, based on
the hadronic $\t$-decay data, see Ref.~\cite{BGMP16}.}

\begin{figure}[t]
\vspace*{4ex}
\begin{center}
\includegraphics*[width=12cm]{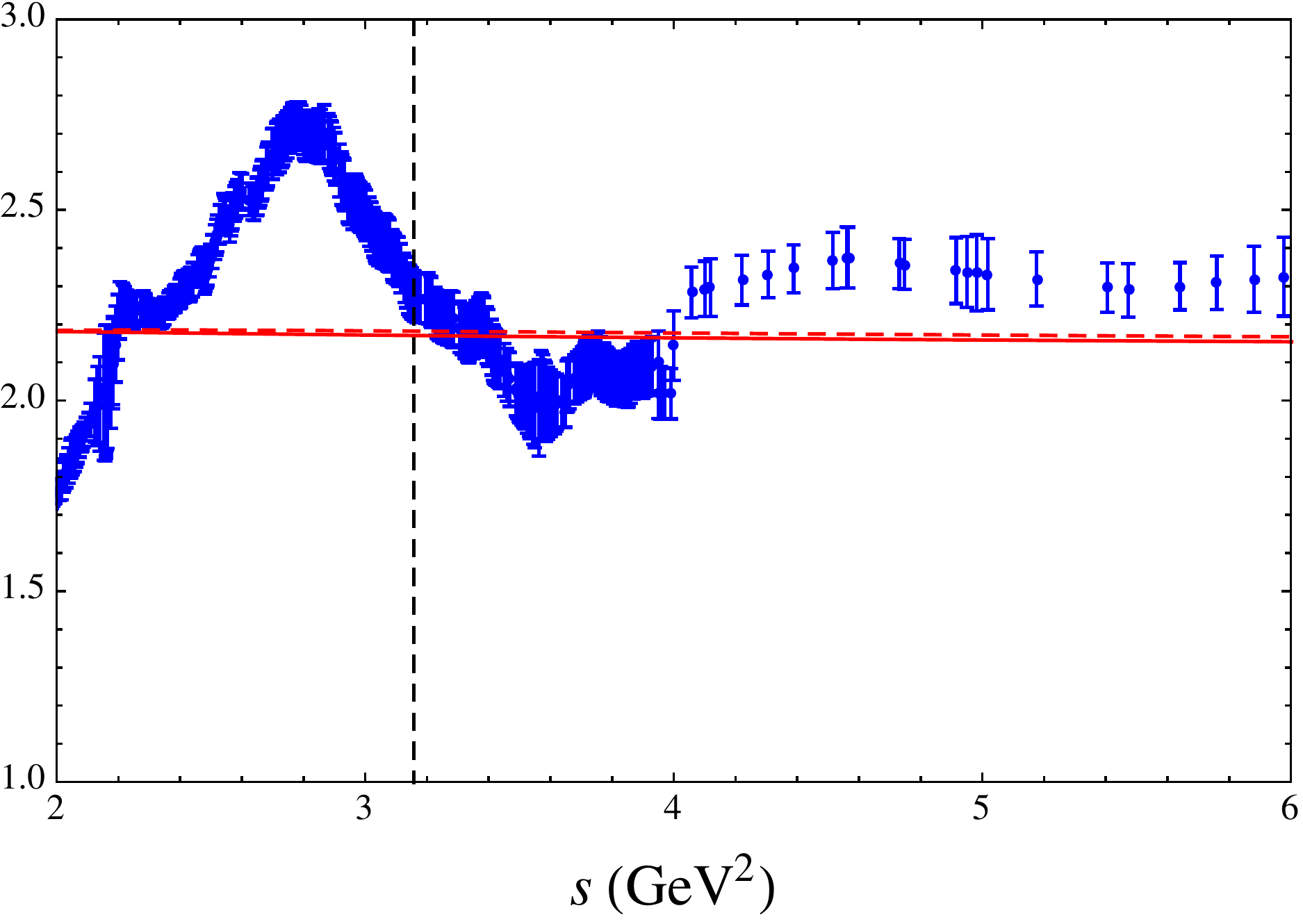}
\end{center}
\begin{quotation}
\floatcaption{Rblowup}%
{{\it A blow-up of $R$-ratio in the region $2\le s\le 6$~{\rm GeV}$^2$
{\rm \cite{KNT18}}.
The red solid and red dashed lines show the results obtained
from perturbation theory with $\a_s(m_\t^2)=0.28$ and $\a_s(m_\t^2)=0.32$,
respectively.   The vertical dashed
line is $s=m_\t^2$.}}
\end{quotation}
\vspace*{-4ex}
\end{figure}

As already mentioned, the aim of this paper is to investigate the validity
of the tOPE strategy using data for the EM spectral function obtained from
$e^+e^-\to\mbox{hadrons}(\g)$. This is not an exercise of academic
interest since, as was shown in Ref.~\cite{BGMP16}, the assumptions
underpinning the tOPE strategy affect the value extracted for $\a_s$ using
this strategy. We will use the recent compilation of exclusive
experimental data for the $R$-ratio provided in Ref.~\cite{KNT18},
which was also recently employed in a determination of $\a_s$ using a
different strategy \cite{alphasEM}. The sum-of-exclusive-modes part of the
compilation of Ref.~\cite{KNT18} reaches up to $s=4$~GeV$^2$, after
which the compilation relies on inclusive data sets. We show
these data in Fig.~\ref{Rblowup} in the region
$2$~GeV$^2\le s\le 6$~GeV$^2$, with the
transition from exclusive to inclusive regions clearly visible.
In Ref.~\cite{alphasEM} it was found that including the much more scarce
data above $s=4$~GeV$^2$ leads to values of $\a_s$ consistent
with those found from the exclusive data, but without decreasing the
error.  Also, using Eq.~(\ref{basicfesr}) with only values of $s_0$ in the inclusive region
leads to a still consistent,
but higher value of $\a_s$ with a much larger error. The reason is
that the inclusive data for the $R$-ratio tends to be larger than what one
would expect from perturbation theory, \seef\ the red solid and dashed
curves in Fig.~\ref{Rblowup}. However, despite the visually apparent
tension between the inclusive data and perturbation theory, it was found
that the larger values of these data are consistent with it being a
statistical fluctuation, given the strong correlations that exist
between the inclusive data points at different values of $s$.

Given all this, we will limit ourselves in this paper to an
investigation of the tOPE approach using the $R$-ratio data up
to $s=4$~GeV$^2$; this is the same region employed in
the determination of $\a_s$ in Ref.~\cite{alphasEM}.

The rest of this paper is organized as follows. Having already
reviewed the tOPE strategy and the goal of this paper
in this section, we discuss in more detail the assumptions on which our 
investigation will rely, and elaborate further on our methodology, in
Sec.~\ref{method}.  Then, in Sec.~\ref{results},
we will present our results, which are shown in Figs.~\ref{optimaldiag}
to~\ref{classicalcorrdoublediff}, and explained in the main text.
A final section restates our assumptions and contains our conclusions.
A preliminary account of this work was presented in Ref.~\cite{prelim}.

\section{\label{method} Assumptions and methodology }
The tOPE strategy has often been employed previously in analyses of FESRs 
based on hadronic $\t$-decay data. As our aim is to use values of $s_0$ 
greater than $m_\tau^2$ to test the strategy, we will focus instead on FESRs
based on EM spectral data.  The spectral function obtained from the $R$-ratio is,
of course, not the same as the spectral functions obtained from $\t$ decays.
Here, we discuss the differences in some detail, spelling out the assumptions
underlying our use of analyses based on the former to cast light on
those based on the latter.

It has been advocated, in the literature applying the tOPE strategy
to $\t$-decay data, that analyses of the sum $V+A$ of the non-strange
$V$ and $A$ channels are preferable, based on the notion that this
sum will be less sensitive to duality violations, and that, in general,
non-perturbative effects may be smaller for the sum than for the
individual $V$ or $A$ channels. Of course, in the EM case
only a $V$-channel spectral function is available, as the photon does
not couple to axial currents. There are two reasons to believe that,
nevertheless, it is reasonable to expect that useful lessons can be
learned by considering the purely $V$-channel EM current  only.

First, it was found in Ref.~\cite{alphasEM} that for a determination of $\a_s$ from
$R$-ratio data in the region above about $3.25$~GeV$^2$ duality violations
can be neglected, with results that are fully consistent with a sum-rule
analysis of the $\t$-based spectral data which modeled duality violations
below the $\t$ mass. This observation implies that the OPE should provide
a good representation of the contour integral over $\P(s)$
in Eq.~(\ref{basicfesrdefsb}) if the radius $s_0$ is chosen to be not
smaller than approximately $3.25$~GeV$^2$. As we will draw our main conclusions
from fits of electroproduction data with values of $s_0$ above the $\t$ mass,
it therefore appears that the issue of sensitivity to duality violations
does not constitute a problem for tests of the tOPE strategy based
on EM $V$-channel data.

Moreover, the tOPE-based results obtained in Ref.~\cite{Pich} show excellent
consistency for the values of $\a_s$ extracted from $V$-channel fits and
$V+A$-channel fits, employing the weights of Eqs.~(\ref{spectralweights})
and~(\ref{optimalweights}). The values obtained from $k\ell$ spectral
weights differ by slightly more than $1\ \s$ between $V$ and
$V+A$,\footnote{Very similar results were found in Ref.~\cite{ALEPH13}.}
while those obtained from optimal weights differ by much less than $1\ \s$.
In both cases, the quality of the $V$-channel fits is better than that of
the $V+A$ channel fits, as measured by the $\c^2$ value per degree of freedom.

Therefore, while we have to assume that $V$-channel-only investigations
can shed light on the tOPE strategy as applied to $\t$ decays, it appears
to us that this is, in fact, a rather innocuous assumption.

A second difference between a $\t$-based analysis and an $R$-ratio-based
analysis is that the non-strange spectral functions obtained from $\t$
decays have isospin $I=1$, while the spectral function obtained from the
$R$-ratio has both $I=1$ and $I=0$ components.\footnote{Since the up and
down quark masses can be taken to vanish in a sum rule extraction of
$\a_s$, we can assume the $\t$-based spectral functions to be purely
$I=1$.} We thus need to assume that the OPE behavior of the scalar polarization
$\P_{\rm EM}$ is similar to that of the scalar polarization $\P_{I=1}$, if we
want to use $R$-ratio based tests to investigate the validity of the
tOPE strategy as applied to analyses of hadronic $\t$-decay data.

In this paper, we will make this assumption, believing that it is well
motivated. If the strange-quark mass $m_s$ were to be negligibly small,
like the up- and down-quark masses,
the $I=1$ and $I=0$ currents would be components of an $SU(3)$-flavor
multiplet, and any conclusions reached in the study of the EM polarization
would directly apply to the $I=1$ case.  For $|s|=s_0$, the OPE for
$\P_{\rm EM}(s)$ differs from that for $\P_{I=1}(s)$ by terms of order
$m_s^2/s_0$, which for $s_0\ge m_\t^2$ is smaller than $m_s^2/m_\t^2\sim
0.003$. Indeed, in Ref.~\cite{alphasEM} it was found that the effect of $m_s$
on the central values of $\a_s(m_\t^2)$ and the OPE condensates $C_{6,8,10}$
is significantly smaller than the fit error on the values of these parameters.
In addition, we emphasize again that Ref.~\cite{alphasEM} finds excellent
agreement between $\a_s$ determinations based on $R$-ratio and the $\t$ data, if
the OPE is treated consistently and the same strategy is used in both cases \cite{alphas1,alphas2,alphas14}.
We conclude that it seems unlikely that the
presence of an $I=0$ component in the EM case would have a significant impact
on the applicability of the tOPE strategy to $R$-ratio data, in
comparison with $\tau$-decay data, at least if the value of $s_0$ is large enough.

In our study of the tOPE, we will repeat the $V$-channel fits carried
out in Ref.~\cite{Pich}, employing the weights~(\ref{spectralweights})
and~(\ref{optimalweights}), but now using the $R$-ratio data compilation
of Ref.~\cite{KNT18}. We will consider values of $s_0$ ranging from $m_\t^2$ to
$4$~GeV$^2$. If we find a good fit, we will compare the experimental
spectral moments $I_w^{\rm exp}(s_0)$ and their theoretical representation
based on that fit $I_w^{\rm th}(s_0)$ as follows. First, if we fit at the
value $s_0=s_0^*$, we compute the differences
\begin{eqnarray}
\label{diffs}
\D_w^{\rm exp}(s_0;s_0^*)&\equiv& I_w^{\rm exp}(s_0)-I_w^{\rm exp}(s_0^*)\ ,\\
\D_w^{\rm th}(s_0;s_0^*)&\equiv& I_w^{\rm th}(s_0)-I_w^{\rm th}(s_0^*)\ ,\nonumber
\end{eqnarray}
as a function of $s_0$. Note that the correlations between spectral
integrals at different $s_0$, as well as those between OPE integrals at
different $s_0$, are very strong; working with the differences in
Eq.~(\ref{diffs}) helps to avoid being misled by these correlations when
comparing experimental and fitted theory integrals. Then, in order to
compare experiment and theoretical representation, we compute
the differences
\begin{equation}
\label{doublediffs}
\D_w^{(2)}(s_0;s_0^*)\equiv\D_w^{\rm th}(s_0;s_0^*)-\D_w^{\rm exp}(s_0;s_0^*)\ ,
\end{equation}
where all correlations, including those between data and fit parameters,
are fully taken into account. Considering these double differences
avoids any issues with under- or over-estimating errors in the comparison
between theory and experiment. Note that, by construction,
$\D_w^{(2)}(s_0^*;s_0^*)=0$, with zero uncertainty.

We will have reason to consider both fully correlated $\c^2$ fits and
what we will refer to as ``diagonal'' fits, where, in the positive
quadratic form to be minimized, we only retain the diagonal part of the
covariance matrix for the integrated spectral data (in computing this
covariance matrix, however, the full data covariance matrix is
taken into account). We emphasize that, when computing errors
on the fitted parameter values produced by such diagonal fits, we take
into account the full covariance matrix for the integrated spectral data,
without ignoring any correlations. Although correlated fits are more
popular, such diagonal fits can also be a useful tool in cases where the
strong correlations make a correlated fit fail. For a detailed explanation
of the diagonal fit procedure, we refer the reader to the
appendix of Ref.~\cite{alphas1}. All correlations, including those
between data and fit parameters, are always fully taken into account in
the computation of the single and double differences
$\D_w^{\rm exp/th}(s_0;s_0^*)$ and $\D_w^{(2)}(s_0;s_0^*)$,
for both types of fits.

\section{\label{results} Results}

\begin{table}[t!]
\begin{center}
\begin{tabular}{|c|c|c|c||c|}
\hline
$s_0^*$ (GeV$^2$) &  $\c^2/$dof  & $p$-value
& $\a_s(m_\t^2)$ & $\a_s(m_\t^2)$ (diag) \\
\hline
$m_\t^2$ & 62.7/1 & $2\times 10^{-15}$  & 0.308(4) & 0.245(10)\\
3.6 & 0.669/1 & 0.41   & 0.264(5) & 0.256(12)\\
\hline
\end{tabular}
\end{center}
\floatcaption{tab1}{\it Fit results with optimal weights. We show the
fits at $s_0^*=m_\t^2$ and at a value of $s_0^*$ for which the fit has a
$p$-value greater than 10\%. Errors shown are fit errors only.}
\end{table}%

\begin{table}[t]
\vspace{0.5cm}
\begin{center}
\begin{tabular}{|c|c|c|c||c|}
\hline
$s_0^*$ (GeV$^2$) &  $\c^2/$dof  & $p$-value
& $\a_s(m_\t^2)$ & $\a_s(m_\t^2)$ (diag)  \\
\hline
$m_\t^2$ & 87.8/1 & $7\times 10^{-21}$  & 0.322(3) & 0.281(6) \\
3.7 & 1.97/1 & 0.16  & 0.277(5) & 0.268(9) \\
\hline
\end{tabular}
\end{center}
\floatcaption{tab2}{\it Fit results with $k\ell$ spectral weights.
We show the fits at $s_0^*=m_\t^2$ and at a value of $s_0^*$ for
which the fit has a $p$-value greater than 10\%.
Errors shown are fit errors only.}
\end{table}%

We begin with showing and discussing some numerical results from fits
employing the tOPE strategy, using fixed-order perturbation theory
(FOPT).\footnote{Results from contour-improved perturbation theory
(CIPT) \cite{CIPT} are very similar, and we will thus restrict
ourselves to FOPT, for simplicity. For detailed studies comparing FOPT
with CIPT, see Refs.~\cite{BJ,BBJ12}.} In Table~\ref{tab1} we show tOPE fit
results with optimal weights. We first attempted a correlated fit at
$s_0^*=m_\t^2$, precisely following the strategy of Ref.~\cite{Pich},
but employing $R$-ratio data instead of hadronic $\t$-decay data. We find,
as the table shows in the first line, that this fit is very bad.
To the right of the double vertical line, we show the corresponding
value of $\a_s(m_\t^2)$ obtained from a diagonal fit. Not surprisingly,
the fit values of $\a_s(m_\t^2)$ for these two fits do not agree. We observe
that, while the correlated fit produces what, nominally at least,
looks like a reasonable result for $\a_s(m_\t^2)$, this result cannot
be accepted because of the very bad fit quality. The value of
$\a_s(m_\t^2)$ from the diagonal fit, on the other hand, is very low
in comparison with the world average, $\a_s(m_\t^2)=0.315(9)$ \cite{PDG,GS17}.

We repeated the same exercise employing $k\ell$ spectral weights, with the
results shown in the first line of Table~\ref{tab2}. The results look
qualitatively similar to those shown in Table~\ref{tab1}, but they are
not in quantitative agreement.

Clearly, our attempts to apply the tOPE strategy of Ref.~\cite{Pich}
at $s_0^*=m_\t^2$ to the electro-production data lead to disastrous
results, and an obvious question is what causes this to happen. Assuming
that there is no problem with the data (which have been extensively used
in Refs.~\cite{alphasEM,KNT18}) leads to the conclusion that the tOPE strategy
does not provide a good fit of the $R$-ratio data, while, according to
Refs.~\cite{ALEPH13,Pich}, it does provide a good fit of the $\t$-decay
data.\footnote{Reference~\cite{BGMP16} confirms this, even though that
reference explains why this does not imply that the tOPE is a reliable
strategy.}  In fact, it was already observed in Ref.~\cite{alphasEM} 
that the OPE does not give a good representation of the $w_\tau$-spectral 
integral of the $R$-ratio data for $s_0\lesssim 3.25\, \mathrm{GeV}^2$ even 
when no terms from the OPE selected by Eq.~(\ref{higherDfesrcont}) were neglected 
in the analysis. Since consistently good fits were obtained at higher values, 
$3.25  \lesssim s_0 \leq 4\, \mathrm{GeV}^2$, in Ref.~\cite{alphasEM}, it seems reasonable 
to infer that $s_0=m_\t^2$ is too small for the OPE to reliably describe the $R$-ratio 
through the spectral integrals appearing in the FESRs~(\ref{basicfesr}) --- even more so 
if, in addition, the OPE is naively truncated.
Another possible contributor to the difference might be that the $R$-ratio spectral integrals, being more 
precise than their counterparts obtained from $\t$-decay data, provide a more stringent test
of the tOPE strategy.\footnote{The
fact that our investigation uses only $V$-channel data, instead of
$V+A$, is much less likely to explain the difference, given the good
quality fits of the $\t$-based $V$ channel data obtained in Ref.~\cite{Pich}.} 

In order to make progress, given this somewhat inconclusive state of
affairs, we proceed to consider fits of Eq.~(\ref{basicfesr}) using a value
$s_0=s_0^*$ larger than $m_\t^2$. We increase $s_0^*$ (in
steps of $0.1$~GeV$^2$, starting from $s_0^*=3.2$~GeV$^2$) 
until the corresponding correlated
fit produces a $p$-value greater than 10\%. For the optimal and $k\ell$
spectral weight sets, we find this occurs for $s_0^*=3.6$~GeV$^2$
and $s_0^*=3.7$~GeV$^2$, respectively. Both correlated and diagonal
fit results are shown in the second lines of Tables~\ref{tab1}
and\ \ref{tab2}, respectively. We see that the results obtained
from correlated and diagonal fits are in good agreement for both set
of weights. However, the correlated fit values for $\a_s(m_\t^2)$
obtained from the optimal and $k\ell$ spectral weight fits are around $2.5~\s$ 
or more apart.\footnote{Note that these two values
are essentially 100\% correlated.}

\begin{figure}[t!]
\vspace*{4ex}
\begin{center}
\includegraphics*[width=14cm]{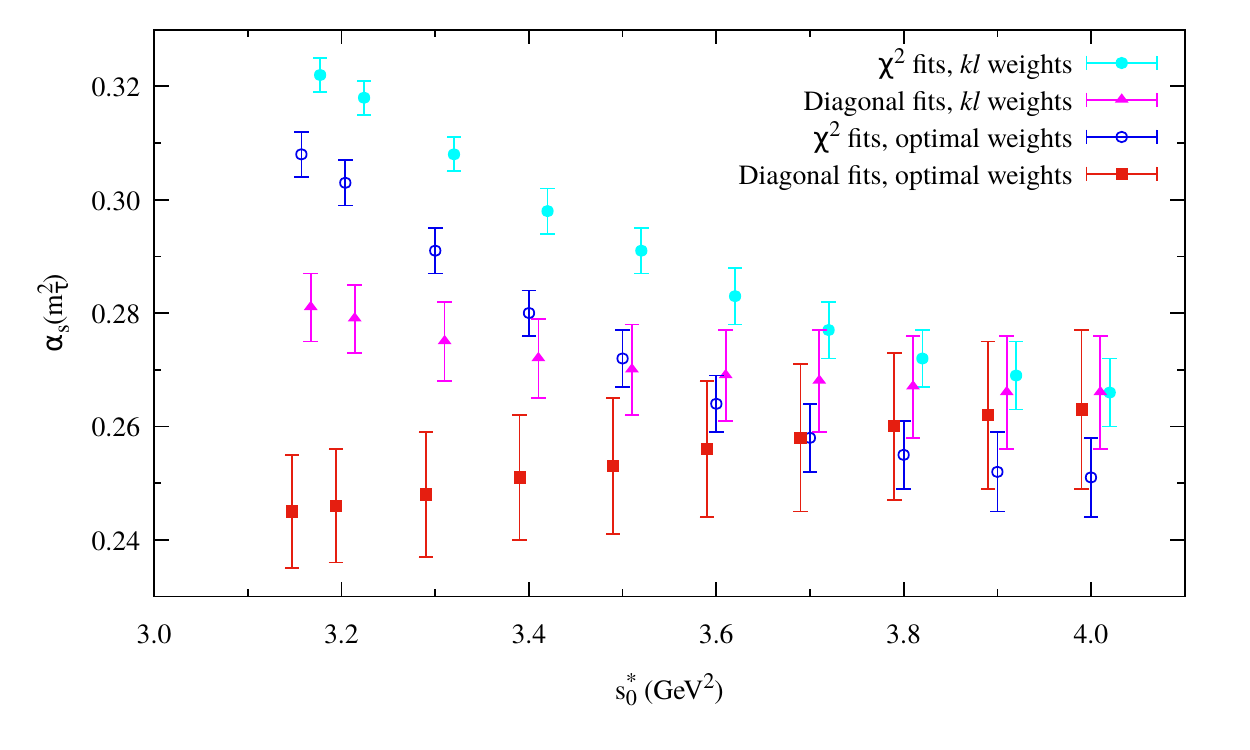}
\end{center}
\begin{quotation}
\floatcaption{alphas}%
{{\it $\a_s(m_\t^2)$ as a function of $s_0$. Blue points (open circles) are
correlated optimal-weight fit results, red  points (squares) are diagonal
optimal-weight fit results, cyan pionts (filled circles) are correlated
$k\ell$-spectral weights fit results, and magenta points (triangles)
are diagonal $k\ell$-spectral weights fit results.  Red, cyan and
magenta points are slightly offset horizontally for better visibility.}}
\end{quotation}
\vspace*{-4ex}
\end{figure}

We show results for the strong coupling for each of these types of fits,
as a function of $s_0^*$, in Fig.~\ref{alphas}, with correlated fit results
for $\a_s(m_\t^2)$ for optimal, respectively, $k\ell$ spectral weights shown
as blue, respectively, cyan points, and diagonal fit results shown as red,
respectively, magenta points. We emphasize that correlated fits with
$s_0^*$ smaller than $3.6$~GeV$^2$ have very small $p$-values, which
rapidly deteriorate down toward $s_0^*=m_\t^2$. In the region where good
correlated fits can be obtained, \ie, for $s_0^*\,\gtap\, 3.6$~GeV$^2$,
there is good agreement with diagonal fits for each set of weights,
with the correlated fits yielding the smaller errors. However, there is
less good agreement between the results obtained using the optimal
and $k\ell$ spectral weight sets. If we were to attempt extracting a
value of $\a_s(m_\t^2)$ from this collection of fits, we would have to
accept a central value of roughly $0.26$--$0.27$, which is again very
low compared to the world average; in particular, compared with the
values obtained in Refs.~\cite{alphas1,alphasEM,ALEPH13,Pich,alphas2,alphas14}.

A much more stringent test of the quality of these fits, and thus
the assumptions underlying the tOPE strategy, is provided by consideration
of the double-differences $\D_w^{(2)}(s_0;s_0^*)$ defined in
Eq.~(\ref{doublediffs}), which we will turn to next. As explained
above, a fundamental assumption of the tOPE strategy is that
it provides a good theoretical description of the data for the spectral
moments $I_w^{\rm exp}(s_0)$ above $s_0\approx m_\t^2$. Given a fit at some
$s_0=s_0^*$, we can vary $s_0$, and plot $\D_w^{(2)}(s_0;s_0^*)$ as a
function of $s_0$. If the assumption is correct, we should find good
agreement between theory (fitted at $s_0^*$) and experiment, for any
value $s_0\ge m_\t^2$.  This means we should find that
$\D_w^{(2)}(s_0;s_0^*)=0$ within errors for $m_\t^2\le s_0\le 4$~GeV$^2$
for all weights $w$ included in the fit.

In Figs.~\ref{optimaldiag} to~\ref{classicalcorrdoublediff} we show
the data and the fitted theory curves, as well as the double differences
$\D_w^{(2)}(s_0;s_0^*)$ for the diagonal fits at $s_0^*=m_\t^2$ in
the optimal (Figs.~\ref{optimaldiag} and~\ref{optimaldiagdoublediff}) and
$k\ell$ spectral (Figs.~\ref{classicaldiag} and~\ref{classicaldiagdoublediff})
weight cases; for the correlated fit at $s_0^*=3.6$~GeV$^2$ in the
optimal weight case (Figs.~\ref{optimalcorr} and~\ref{optimalcorrdoublediff});
and for the correlated fit at $s_0^*=3.7$~GeV$^2$ in the $k\ell$ spectral
weight case (Figs.~\ref{classicalcorr} and~\ref{classicalcorrdoublediff}).
We emphasize again that all errors shown in all five panels
in all four double-difference plots
have been computed taking all correlations, including those between data and
the fit parameters, into account.

Figures~\ref{optimaldiag},~\ref{classicaldiag},~\ref{optimalcorr},
and~\ref{classicalcorr}, which show the data compared with the fitted
theory curves (using central values for the fit parameters) show what
the fits look like, as a function of $s_0$. It is clear that the theory,
fixed by a fit at $s_0=s_0^*$, does not do a very good job of describing
the $s_0$ dependence, but, given the strong correlations, it is hard
to ascertain, from these figures alone, how bad this problem actually
is. It is for this reason that we focus on the comparison between
experiment and theory provided by the ``double-difference'' figures,
where we consider the quantity $\D_w^{(2)}(s_0;s_0^*)$ as a function of
$s_0$ for the fixed values of $s_0^*$ used in the corresponding fits.
These results are shown in Figs.~\ref{optimaldiagdoublediff},
~\ref{classicaldiagdoublediff},~\ref{optimalcorrdoublediff},
and~\ref{classicalcorrdoublediff}.

Three different observations are of relevance to assessing the lessons
to be learned from the results shown in the double-difference figures.
First, in obtaining results from a fit at $s_0=s_0^*$, it has been
assumed that the tOPE strategy provides a valid theory representation
at that value of $s_0$. This implies, with certainty, that this strategy
should provide a good theory representation for any value of
$s_0\ge s_0^*$. The plots of $\D_w^{(2)}(s_0;s_0^*)$ for $s_0>s_0^*$
directly test this assumption. Second, if the claim is that the
tOPE strategy works at $s_0=m_\t^2$, this implies that, for any
$s_0^*\in [m_\t^2, 4~\mbox{GeV}^2]$, $\D_w^{(2)}(s_0;s_0^*)$ should
be consistent with zero for all $s_0\ge m_\t^2$, irrespective of the
value of $s_0^*$ used in the fit. Finally, for a fit to five spectral
moments (whether employing Eq.~(\ref{spectralweights}) or Eq.~(\ref{optimalweights}))
to be successful, $\D_w^{(2)}(s_0;s_0^*)$ has to be consistent with zero
as a function of $s_0$ for each of the five weights in the set. If
$\D_w^{(2)}(s_0;s_0^*)$ shows a significant deviation from zero
for just one or two weights, this indicates a problem with the fit,
and thus with the tOPE strategy.

From Figs.~\ref{optimaldiag},~\ref{classicaldiag},~\ref{optimalcorr},
and~\ref{classicalcorr}, we see that the fitted theory curves do
agree within errors with all five of the corresponding 
weighted spectral integrals for $s_0$ in the vicinity of 
the tOPE fit point $s_0=s_0^*$.  This is, however, not typically
the case for $s_0$ farther away from $s_0^*$.
One clearly observes, however, that,
for many weights, $\D_w^{(2)}(s_0;s_0^*)$, shown in
Figs.~\ref{optimaldiagdoublediff},~\ref{classicaldiagdoublediff},
~\ref{optimalcorrdoublediff}, and~\ref{classicalcorrdoublediff}, is not
consistent with zero for $m_\t^2\le s_0\le 4$~GeV$^2$, and that
many points are, in fact, many $\s$ away from zero. This is particularly
true for the correlated fits shown in Figs.~\ref{optimalcorrdoublediff}
and~\ref{classicalcorrdoublediff} in the region $3.25$~GeV$^2\le s_0<s_0^*$.
This casts serious doubt on the validity of the tOPE strategy in the
whole region we have investigated here, \ie, the region between
$s\approx m_\t^2$ and $s=4$~GeV$^2$.

\section{\label{conclusion} Discussion and conclusion}

In this paper, we have continued our investigation of the validity of the
truncated OPE (tOPE) approach to FESR analyses, an investigation
of relevance, for example, to the determination of $\a_s$ from such analyses
of hadronic $\t$-decay data. The key observation is that if the tOPE
approach works at a ``fit point'' $s_0^*$ near $m_\t^2$, it should (a) certainly
work at higher values of $s_0^*$, and (b) given a fit at $s_0=s_0^*$ equal
to $m_\t^2$ or higher, there should be good agreement between the
experimental spectral moments and the theory representations employing
the OPE parameter fit values at $s_0\ge s_0^*$.

Under rather mild assumptions, tests of these two observations can be
carried out using $R$-ratio data, for which very precise results
are available up to $s=4$~GeV$^2$ \cite{KNT18,DHMZg}. The two assumptions
are that (i) it is sufficient to consider only EM vector-channel
data (with the axial-channel data available in $\t$ decays not being
accessible through $e^+e^-\to\mbox{hadrons}$), and (ii) the presence of an
$I=0$ component in the $R$-ratio data does not change the behavior of the OPE
in an essential way as far as the tOPE strategy is concerned. We
discussed these two assumptions, and the reasons for expecting them to
be reliable, in detail in Sec.~\ref{method}.

The tOPE approach makes two basic assumptions. The first is that
violations of quark-hadron duality can be ignored already at energies as
low as the $\t$ mass, and the second that the expansion in $1/s_0$
of the integrated OPE, though in actual fact divergent, acts as if it
were rapidly converging already at $s_0=m_\t^2$. If we assume that in
the region $s_0\,\gtap\,m_\t^2$ duality violations are relatively
unimportant (an assumption that is consistent with the results of
Ref.~\cite{alphasEM}),\footnote{The fact that the
weights~(\ref{spectralweights}) and~(\ref{optimalweights}) are doubly
pinched, also serves to suppress such integrated duality violations.}
the question centers on the nature of the OPE for values of
$Q^2$ with $\vert Q^2\vert = s_0$ in this region.

Our central results are shown in Figs.~\ref{optimaldiagdoublediff},~\ref{classicaldiagdoublediff},~\ref{optimalcorrdoublediff},
and~\ref{classicalcorrdoublediff}. If the tOPE were to be valid,
they should show data points consistent with zero for all $s_0\ge m_\t^2$.
Instead, these figures show very significant disagreements,
as a function of $s_0$, between
the experimental values of the spectral moments and the theory
representations based on the tOPE fits at various fit points $s_0^*$,
for both the optimal-weight- and $k\ell$-spectral-weight-based fits.
We emphasize that the error bars shown for the double differences,
defined in Eq.~(\ref{doublediffs}), take all correlations into
account, including those between data and the fit parameter values.

The first two of these figures show the results of diagonal fits
at $s_0^*=m_\t^2$, which we considered after we found that
correlated fits do not work
at this $s_0^*$ (\seef\ Tables~\ref{tab1} and~\ref{tab2}). More important
are the fits at $s_0^*=3.6$~GeV$^2$ (for optimal weights, shown in
Fig.~\ref{optimalcorrdoublediff}) and $s_0^*=3.7$~GeV$^2$ (for $k\ell$
spectral weights, shown in Fig.~\ref{classicalcorrdoublediff}).
These are correlated $\c^2$ fits with acceptable $p$-values, where
the figures nonetheless show very serious mismatches between the
experimental data and the theory representations provided by the
fits. We observe that if the tOPE strategy does not work
for a value of $s_0^*$ significantly larger than $m_\t^2$, it
certainly cannot be expected to be reliable at $s_0^*=m_\t^2$,
thus making our tests at $s_0^*=m_\t^2$ less relevant. Nevertheless,
the fact that correlated fits at $s_0^*=m_\t^2$ do not work stands
in sharp contrast to what is found with data from hadronic $\t$
decays \cite{BGMP16,Pich}. 

We may also ask what our results imply for the OPE itself, rather than just
for the tOPE strategy. Our tests probe the integrated OPE up to
dimension 16, for both the optimal and $k\ell$ spectral weight sets. As the
OPE is (at best) an asymptotic expansion, the question is to which order
one can expect to be able to use it while still having the truncated expansion
approach the underlying
true physical value. The answer to this question will, of course, depend on the
value of $s_0$. The analysis in Ref.\,\cite{alphasEM} showed that the
OPE provides a consistent representation of the EM vacuum polarization
up to dimension $D=10$, a conclusion supported, in particular,
by the consistency of the results for the effective $D=6$ condensate, $C_6$,
obtained using different weights with degree up to $4$, in the region
$m_{\tau}^2\leq s_0\leq 4\, \mathrm{GeV}^2$. It is possible
that the OPE starts to already diverge before one reaches the term of
dimension $16$ for $s_0$ in the range between $m_\t^2$ and 4~GeV$^2$, but
it is also possible that it approaches the (unknown) exact answer
reasonably well, to this order, and in this range. Our tests leave this
question undecided.  What they do show is that even if the OPE is
still approaching the true answer out to $D=16$, it is not doing so
rapidly enough that terms with $D>10$ (for the optimal weights) or $D>8$
(for the $k\ell$ spectral weights) can be neglected, in the sum rules
considered here. In any case, our analysis provides a clear message
that it is safest to restrict the analysis to those observables
which only receive a contribution from the lower-dimension terms in the OPE.

Our main conclusion is that even if one considers an energy region in which
duality violations are likely to be strongly suppressed, the tOPE
strategy leads to inconsistent results, thus invalidating the neglect of
higher-order terms in the OPE in that energy region. Taken together with our
earlier investigations of the tOPE strategy reported in Ref.~\cite{BGMP16},
the implication is that the tOPE strategy is not a reliable one.
It should thus no longer be employed, for example, in the extraction of
$\a_s$ from hadronic $\t$ decays or $R$-ratio data, up to at least
$s=4$~GeV$^2$, particularly since an alternative method which does not
suffer from the shortcomings of the tOPE strategy
exists \cite{alphas1,alphasEM,alphas2,alphas14}.

\vspace{3ex}
\noindent {\bf Acknowledgments}
\vspace{3ex}

DB,  KM and SP would like to thank the Department of Physics and Astronomy at
San Francisco State University for hospitality, and MG would like to thank the
Department of Mathematics and Statistics at York University for hospitality.
The work of DB is supported by the S{\~a}o Paulo Research Foundation
(FAPESP) Grant No. 2015/20689-9 and by CNPq Grant No. 309847/2018-4.
The work of MG is supported by the U.S. Department of Energy, Office of Science,
Office of High Energy Physics, under Award Number DE-FG03-92ER40711.
KM is supported by a grant from the Natural Sciences and Engineering Research
Council of Canada.
SP is supported by CICYTFEDER-FPA2017-86989-P and by Grant No. 2017 SGR 1069.


\newpage
\begin{figure}[h]
\vspace*{4ex}
\begin{center}
\includegraphics*[width=7.4cm]{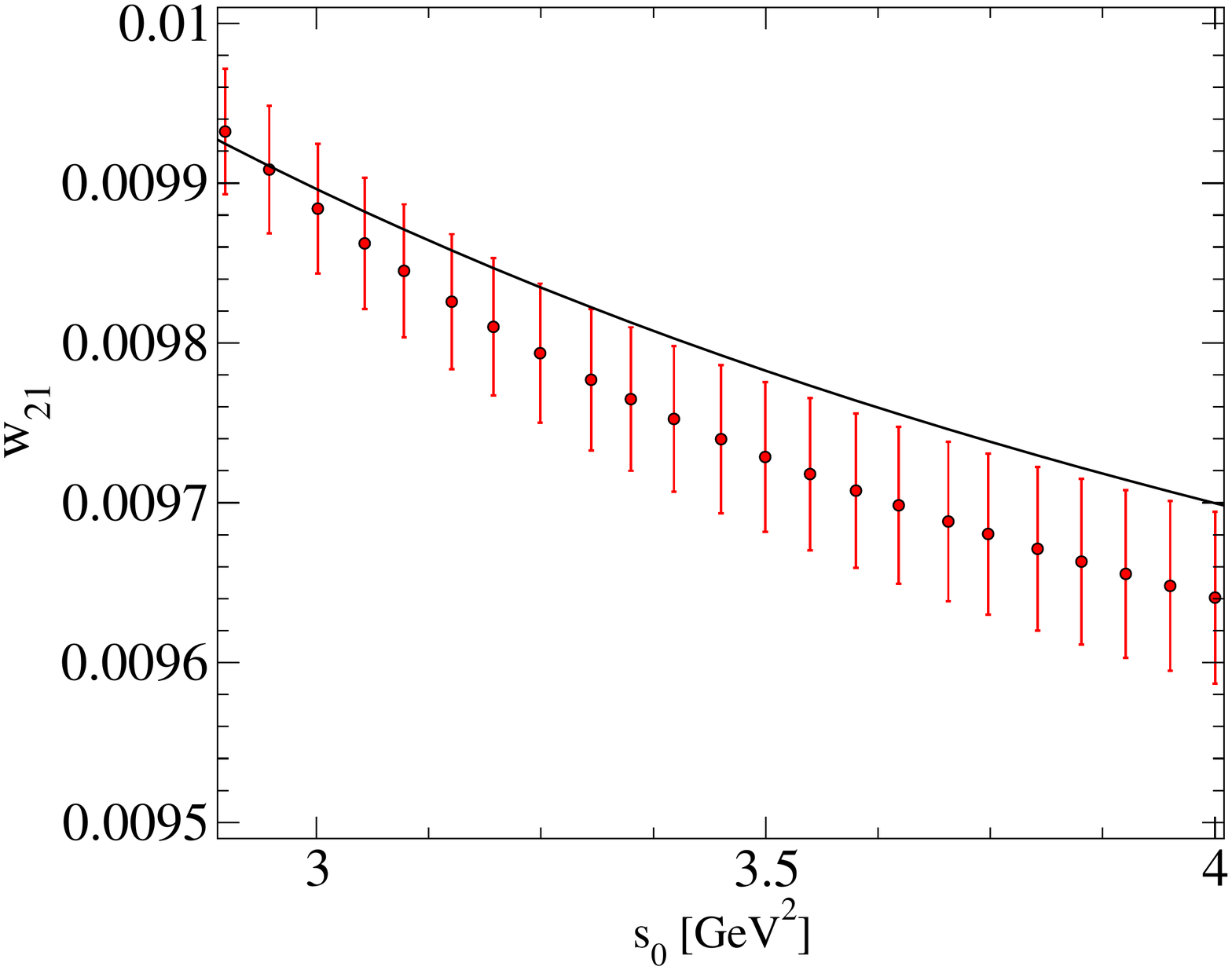}
\hspace{0.1cm}
\includegraphics*[width=7.4cm]{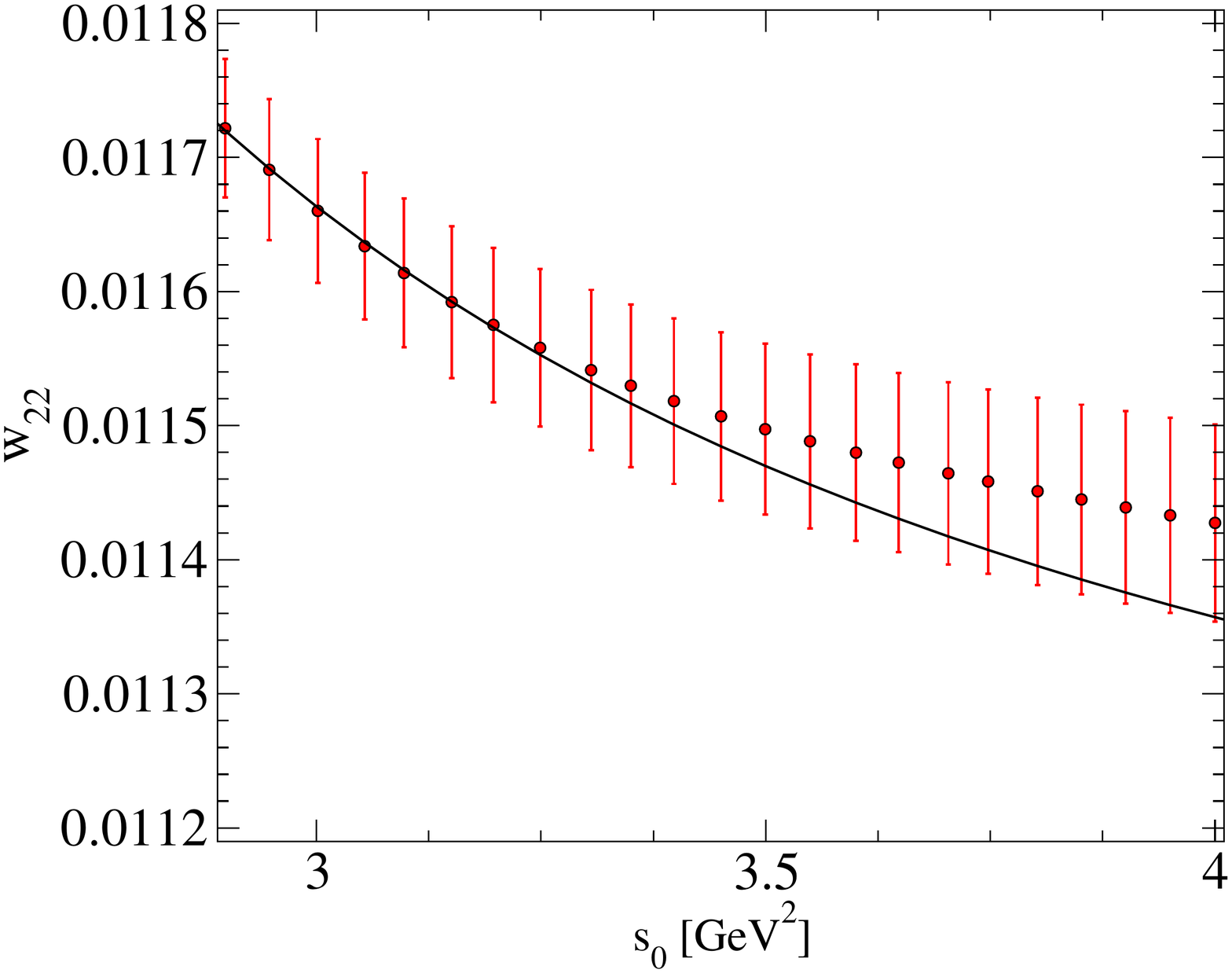}

\vspace{0.5cm}
\includegraphics*[width=7.4cm]{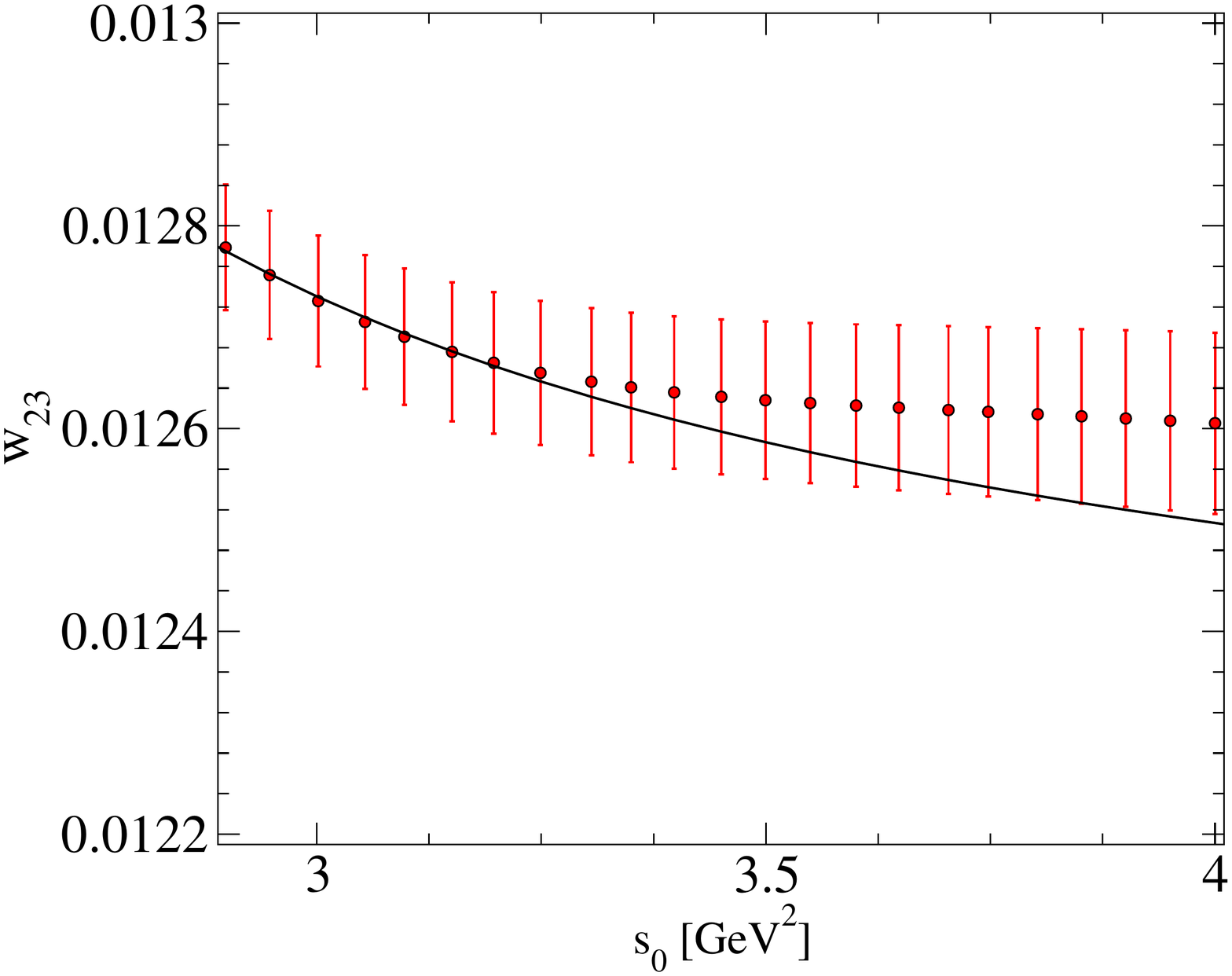}
\hspace{0.1cm}
\includegraphics*[width=7.4cm]{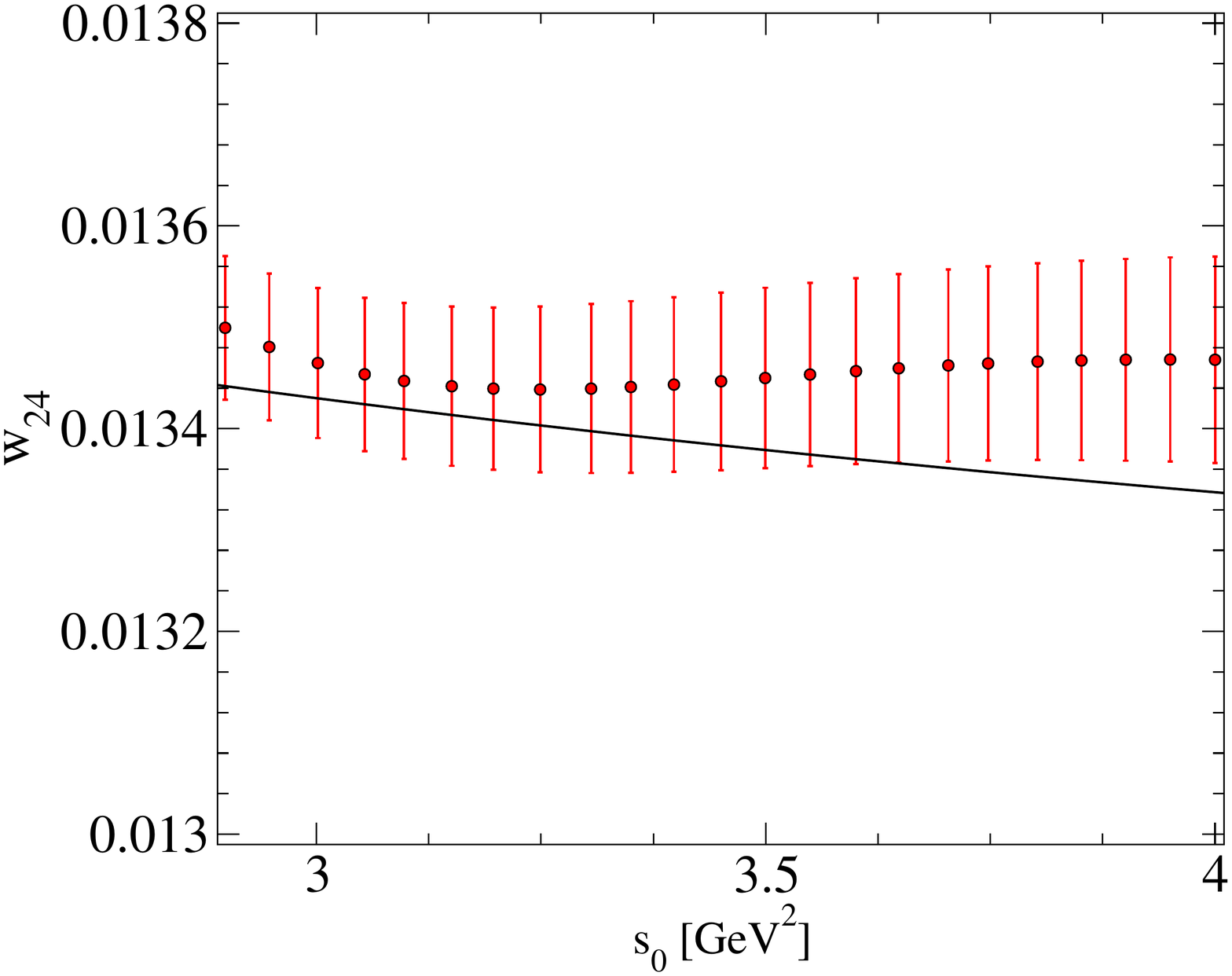}

\vspace{0.5cm}
\includegraphics*[width=7.4cm]{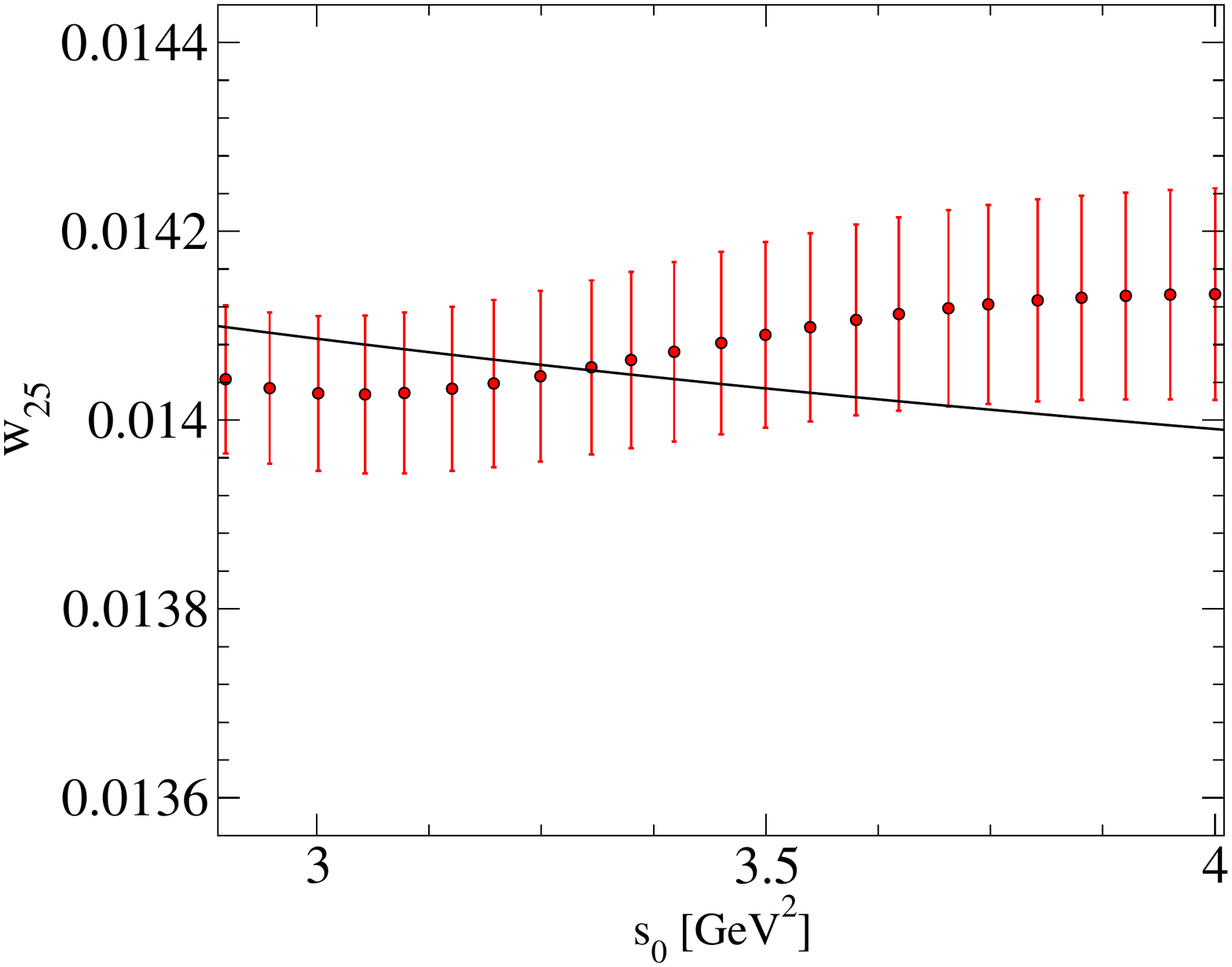}
\end{center}
\begin{quotation}
\floatcaption{optimaldiag}%
{{\it Comparison of $I_w^{\rm exp}(s_0)$ with $I_w^{\rm th}(s_0)$ with parameter
values obtained from diagonal fits with optimal weights, as a function of
$s_0$ with $s_0^*=m_\t^2$.}}
\end{quotation}
\vspace*{-4ex}
\end{figure}

\begin{figure}[t]
\vspace*{4ex}
\begin{center}
\includegraphics*[width=7.4cm]{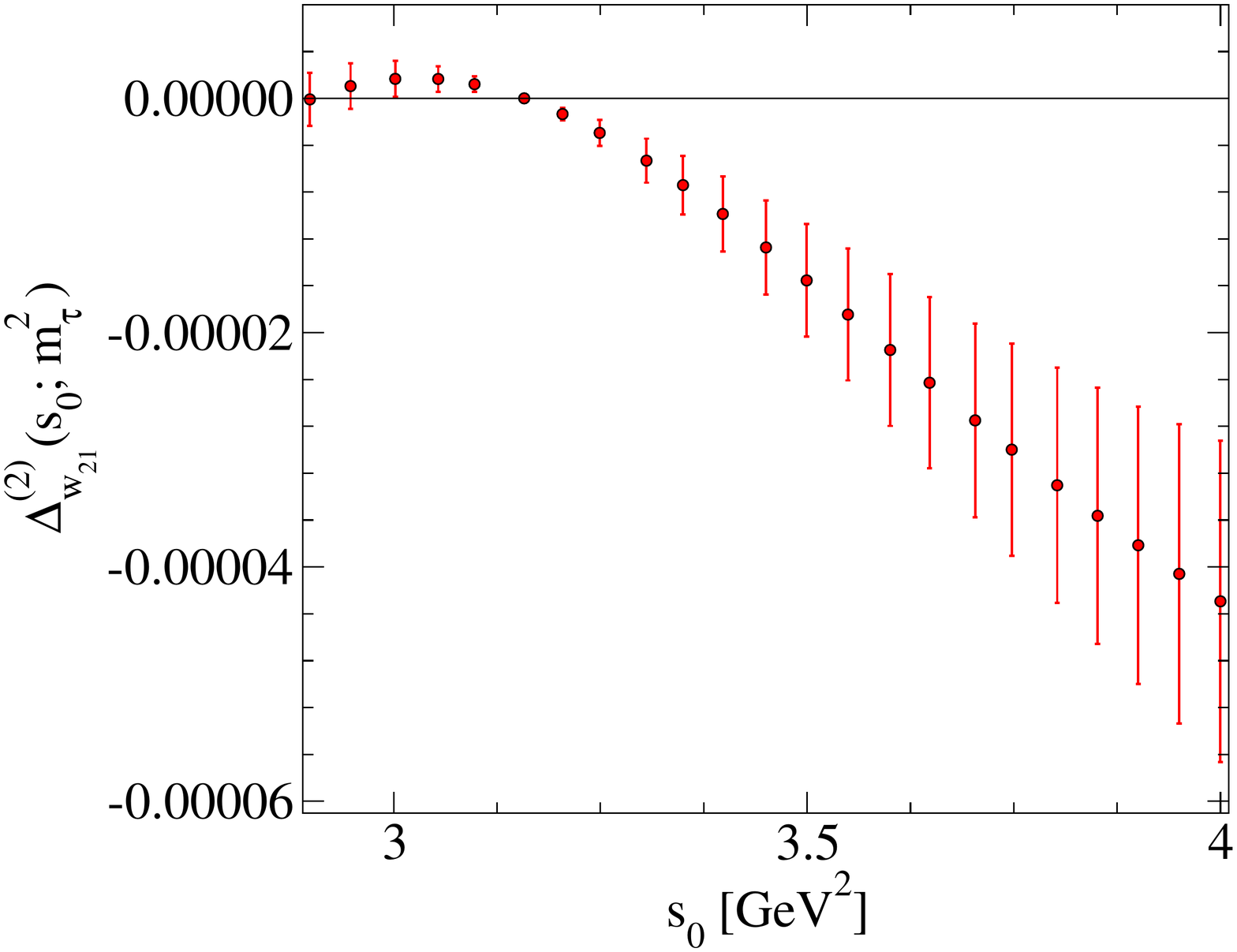}
\hspace{0.1cm}
\includegraphics*[width=7.4cm]{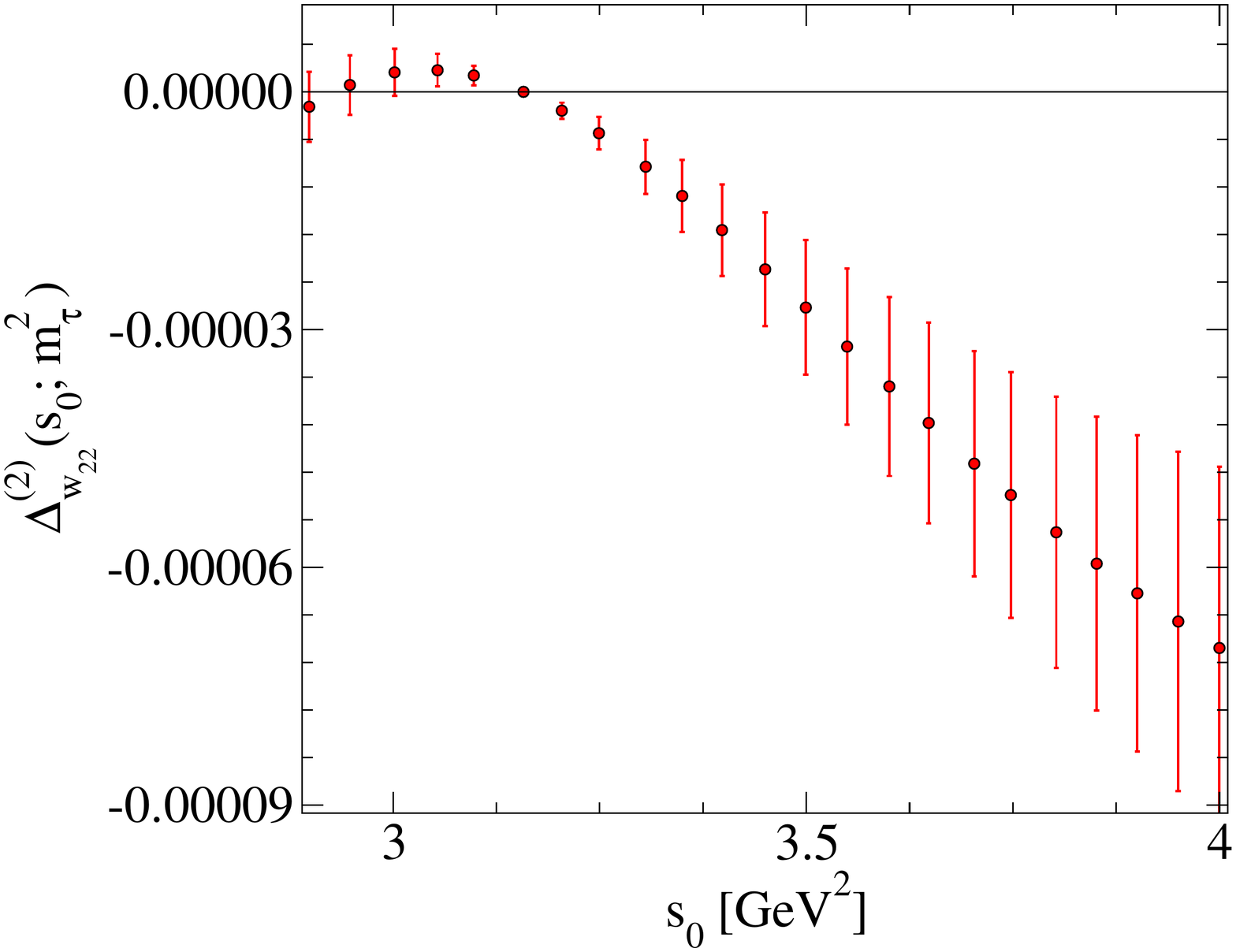}

\vspace{0.5cm}
\includegraphics*[width=7.4cm]{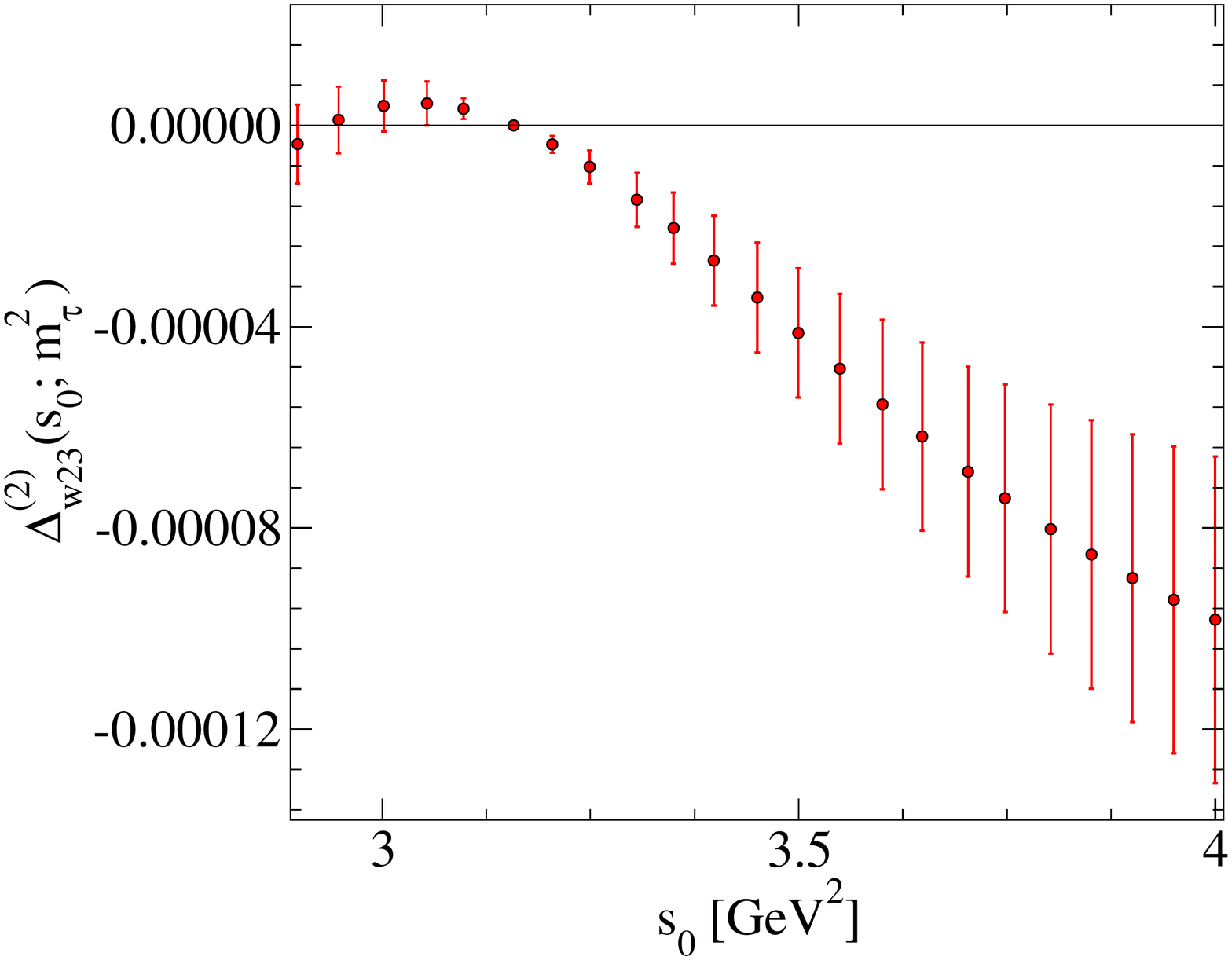}
\hspace{0.1cm}
\includegraphics*[width=7.4cm]{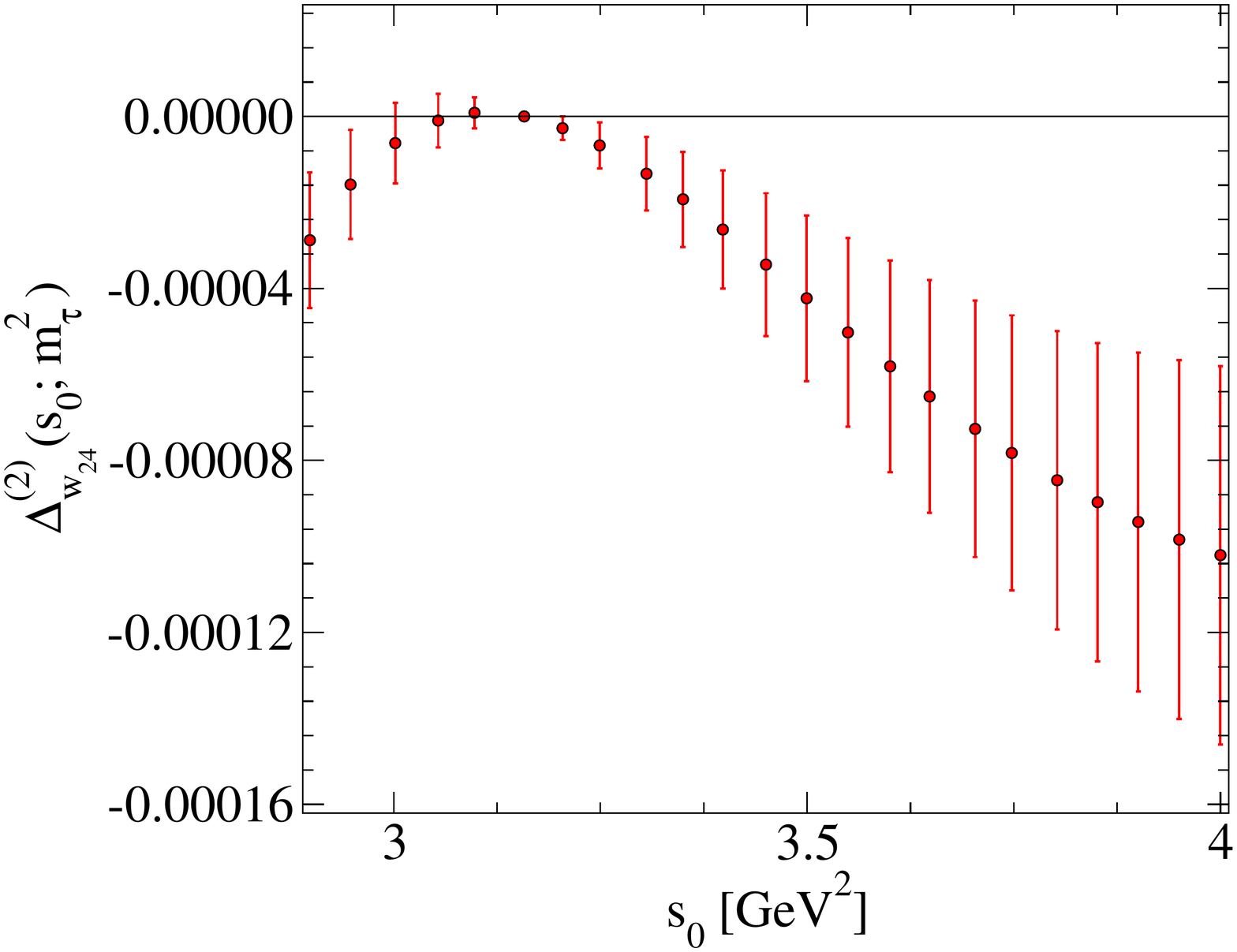}

\vspace{0.5cm}
\includegraphics*[width=7.4cm]{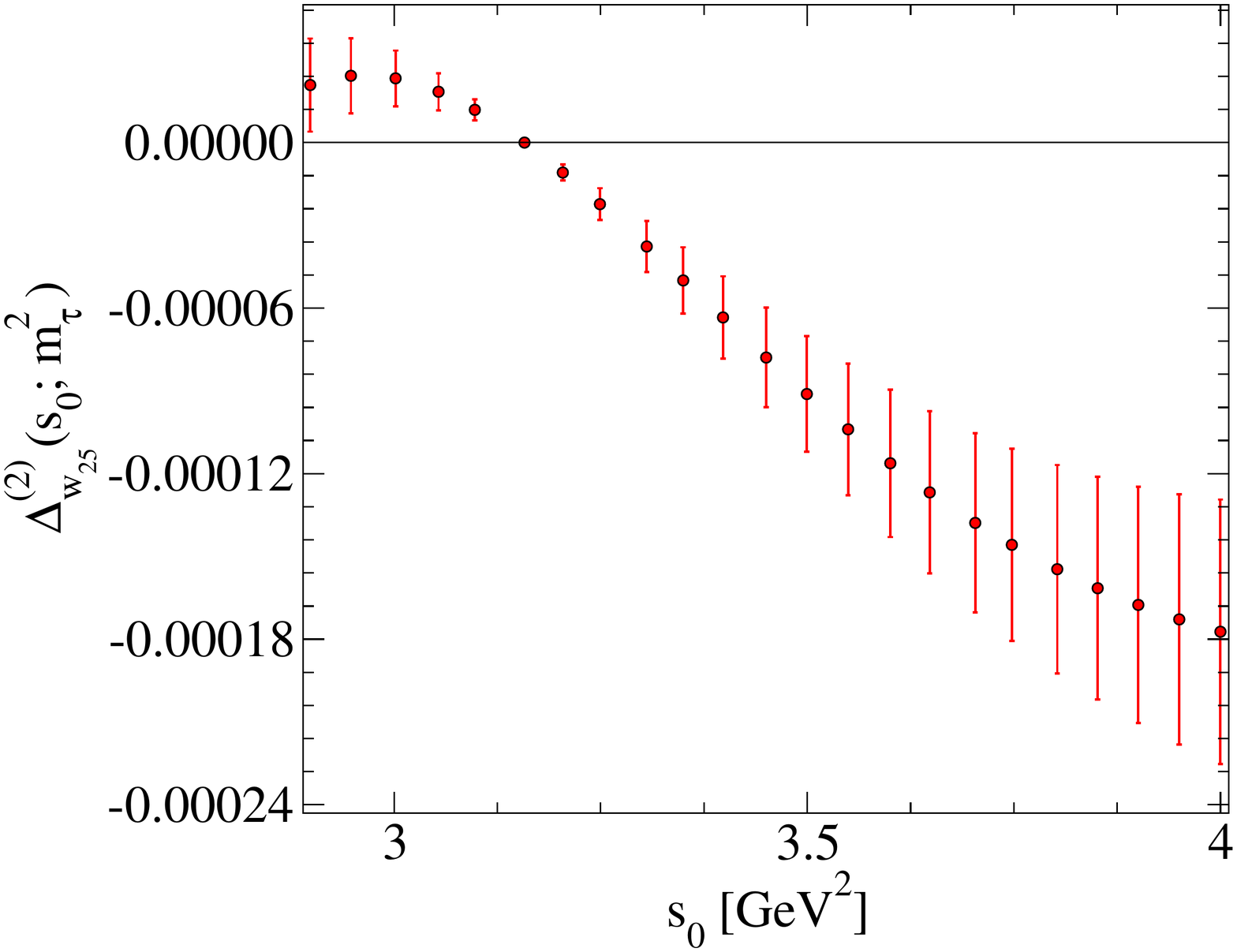}
\end{center}
\begin{quotation}
\floatcaption{optimaldiagdoublediff}%
{{\it The double differences, $\D^{(2)}_w(s_0;s_0^*)$, obtained from
diagonal fits with optimal weights, as a function of $s_0$ with
$s_0^*=m_\t^2$.}}
\end{quotation}
\vspace*{-4ex}
\end{figure}

\begin{figure}[t]
\vspace*{4ex}
\begin{center}
\includegraphics*[width=7.4cm]{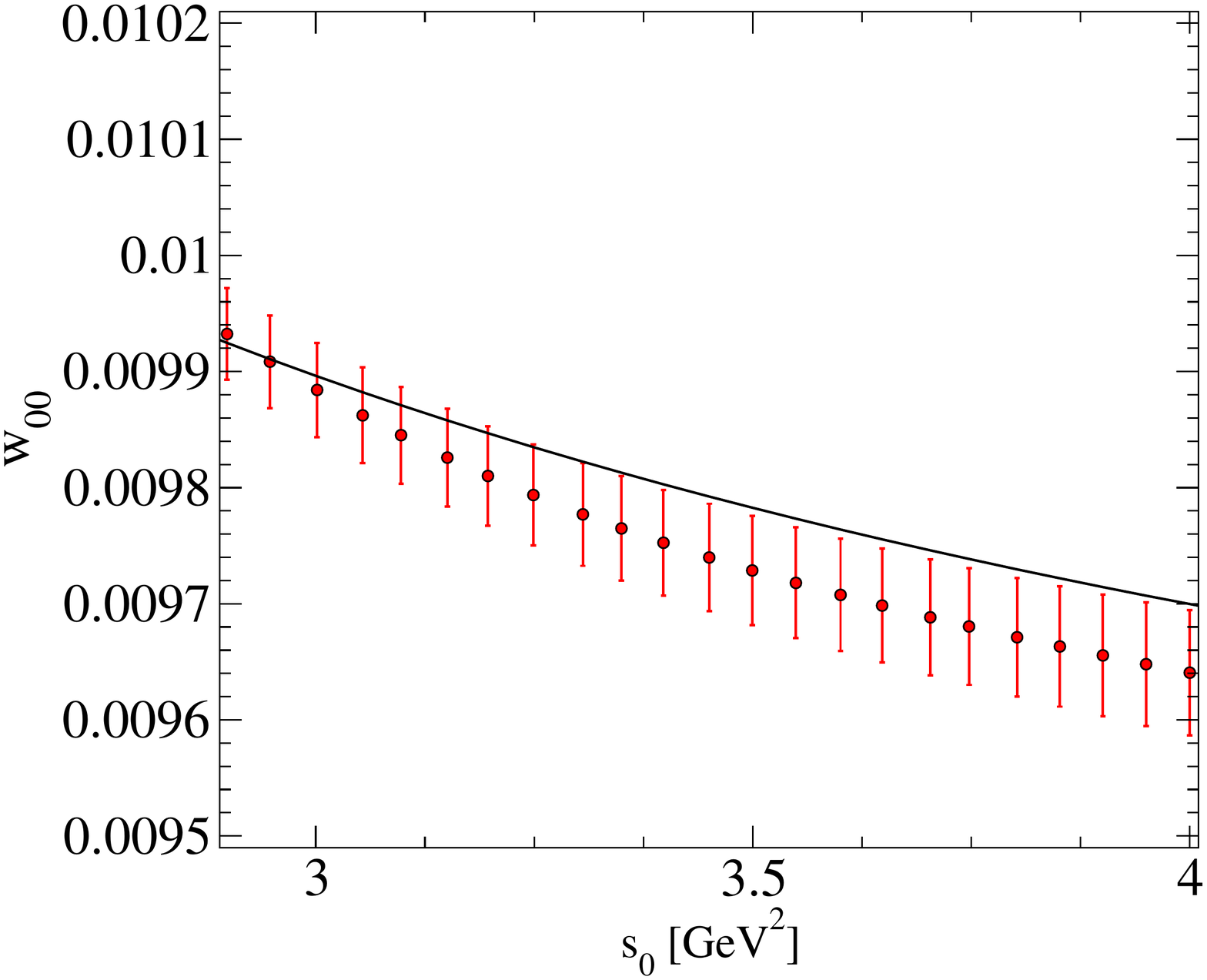}
\hspace{0.1cm}
\includegraphics*[width=7.4cm]{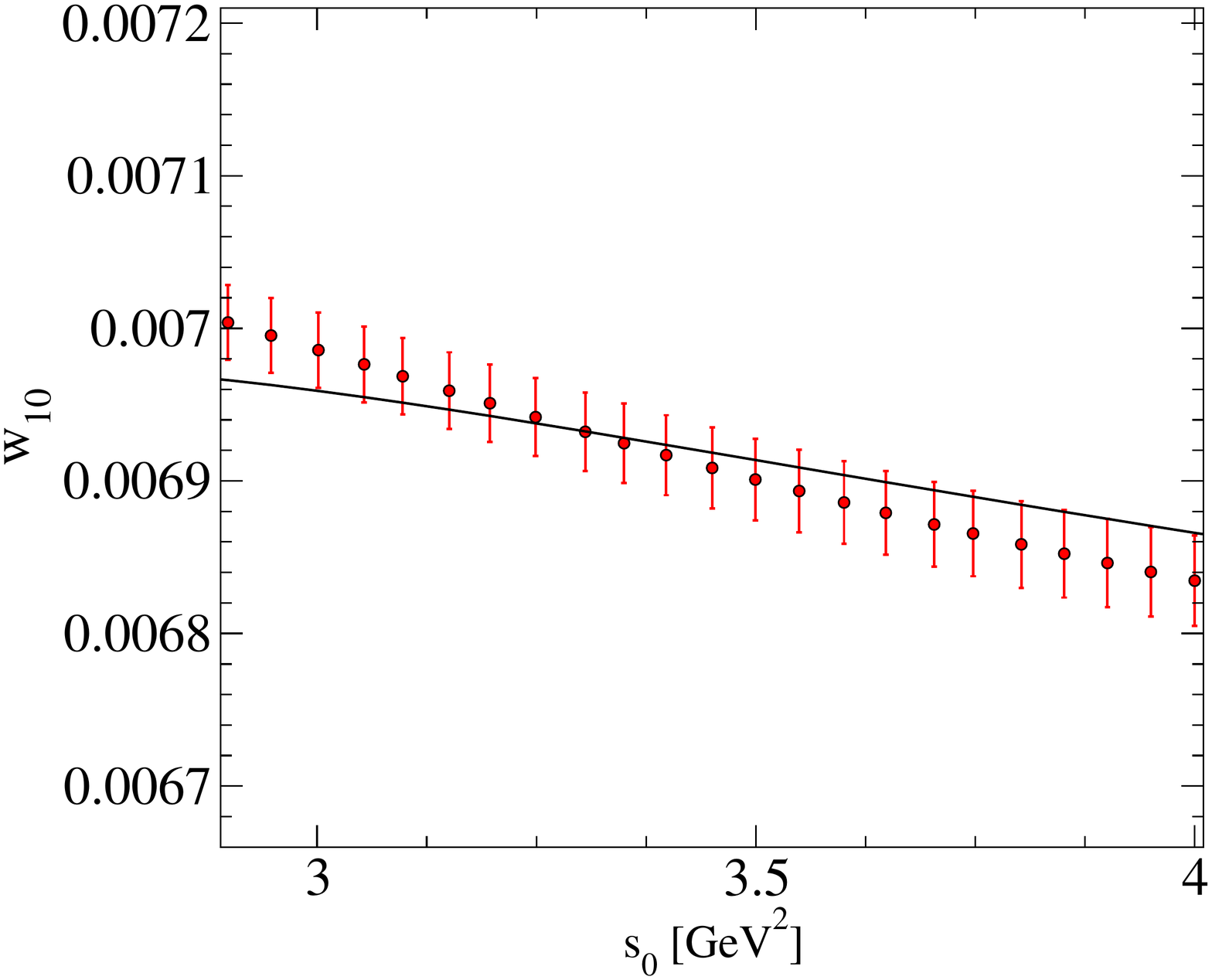}

\vspace{0.5cm}
\includegraphics*[width=7.4cm]{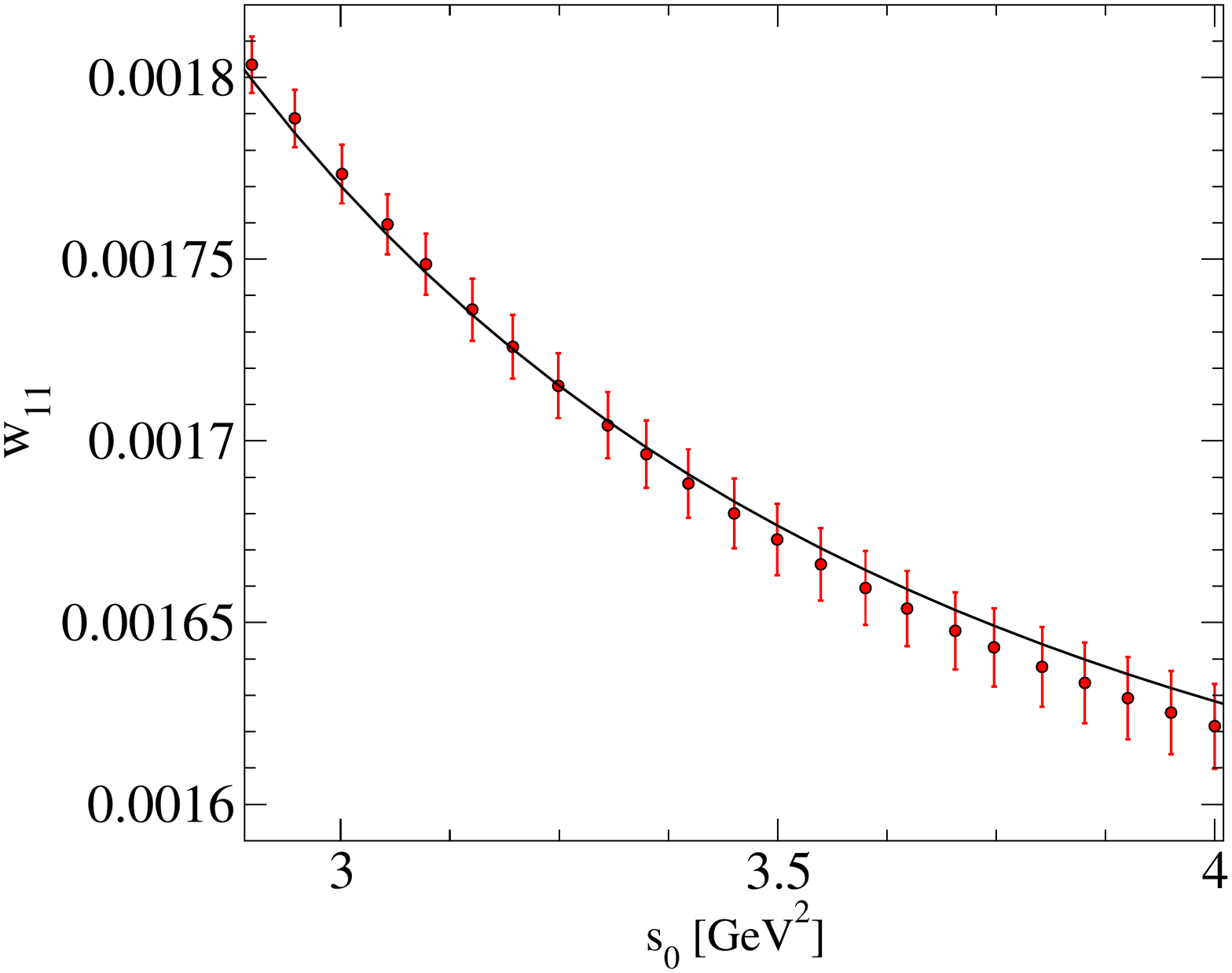}
\hspace{0.1cm}
\includegraphics*[width=7.4cm]{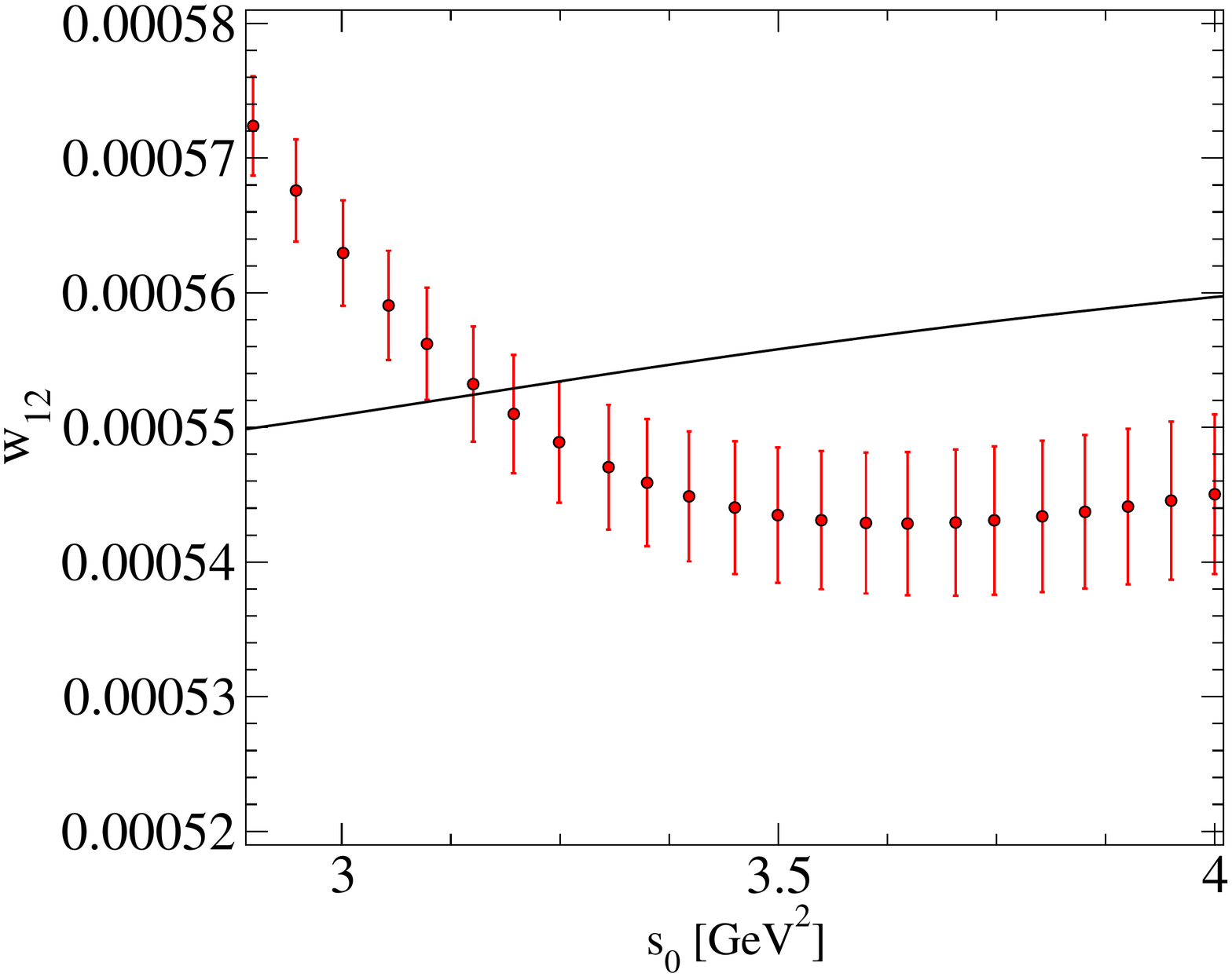}

\vspace{0.5cm}
\includegraphics*[width=7.4cm]{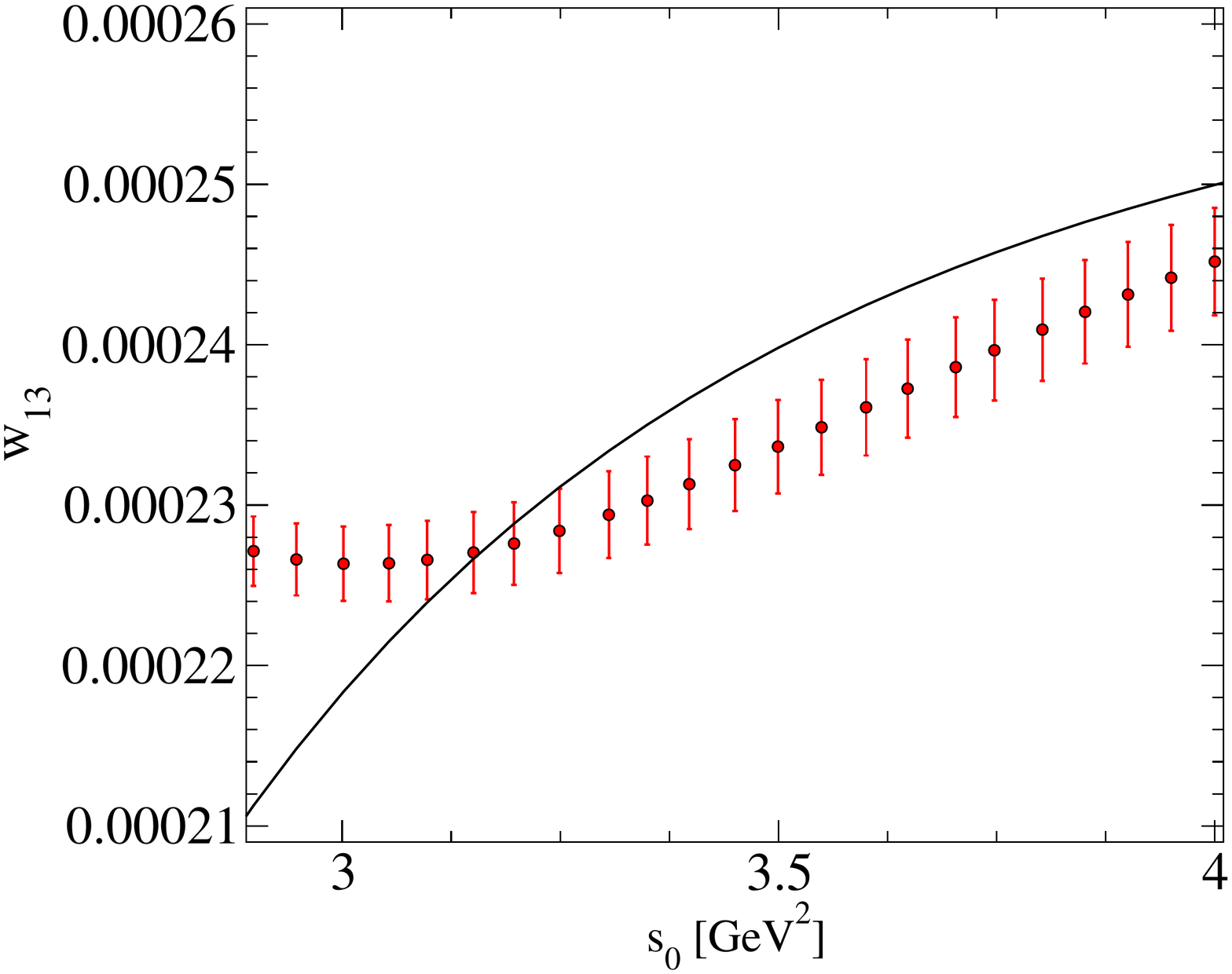}
\end{center}
\begin{quotation}
\floatcaption{classicaldiag}%
{{\it Comparison of $I_w^{\rm exp}(s_0)$ with $I_w^{\rm th}(s_0)$ with parameter
values obtained from diagonal fits with $k\ell$ spectral weights, as a
function of $s_0$ with $s_0^*=m_\t^2$.}}
\end{quotation}
\vspace*{-4ex}
\end{figure}

\begin{figure}[t]
\vspace*{4ex}
\begin{center}
\includegraphics*[width=7.4cm]{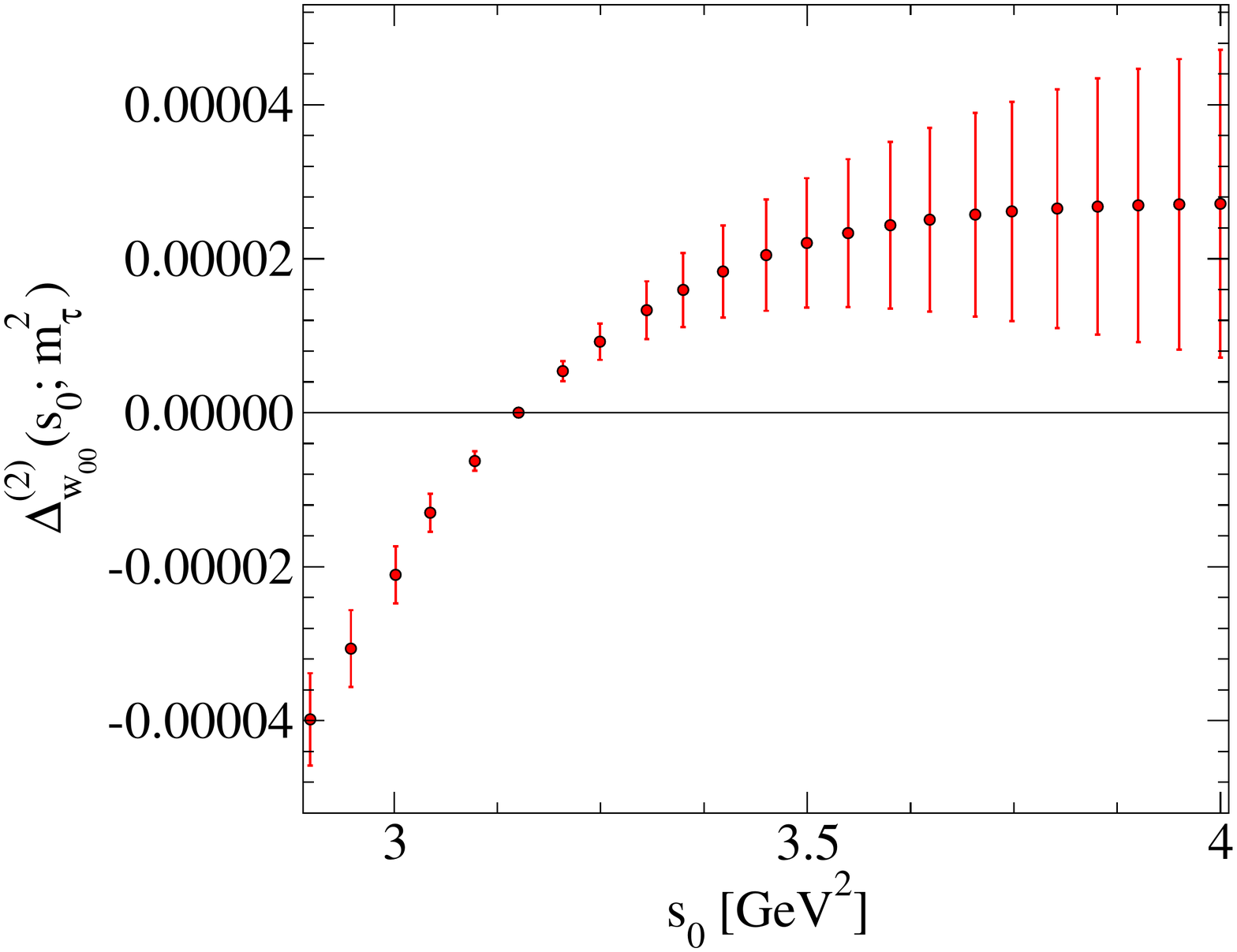}
\hspace{0.1cm}
\includegraphics*[width=7.4cm]{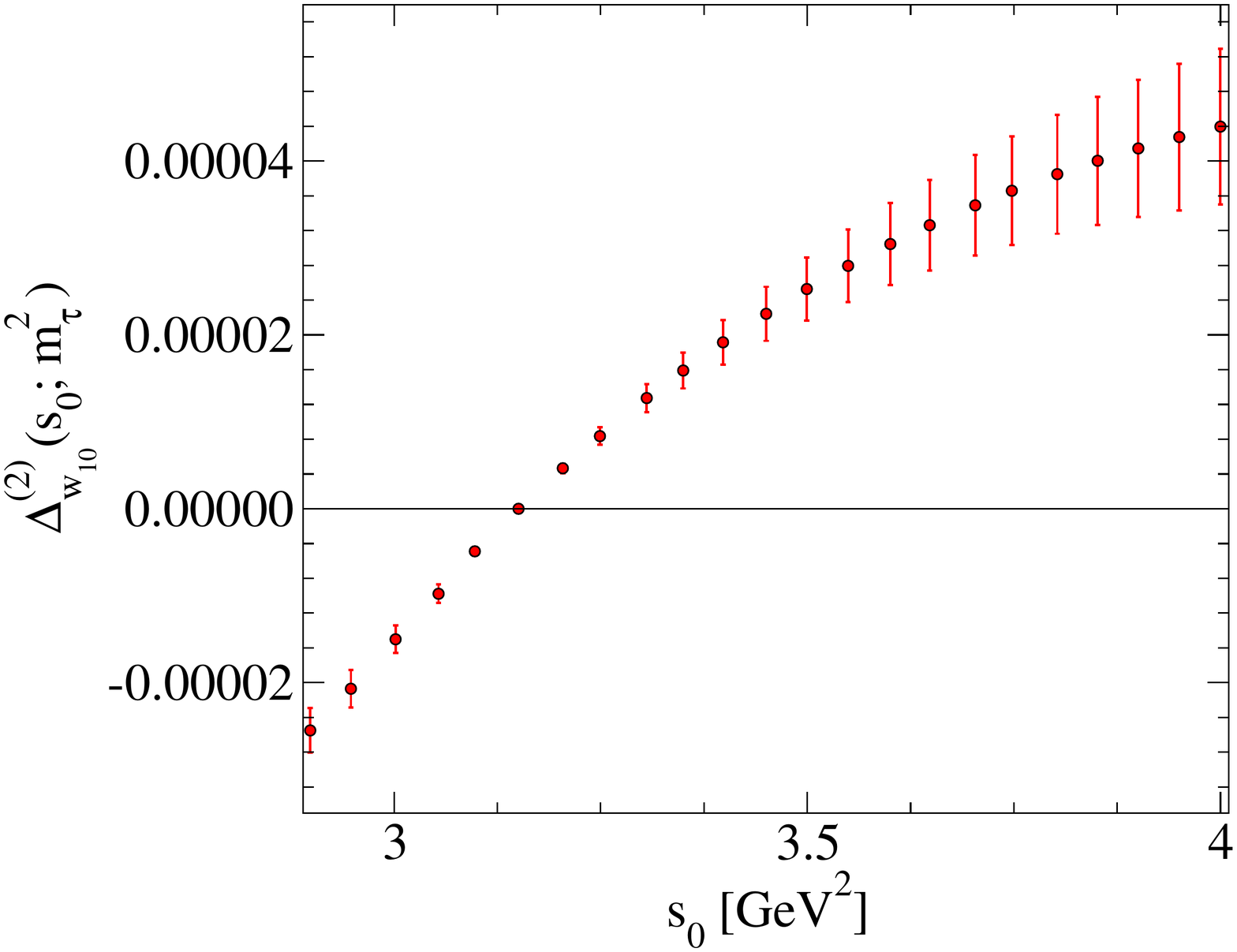}

\vspace{0.5cm}
\includegraphics*[width=7.4cm]{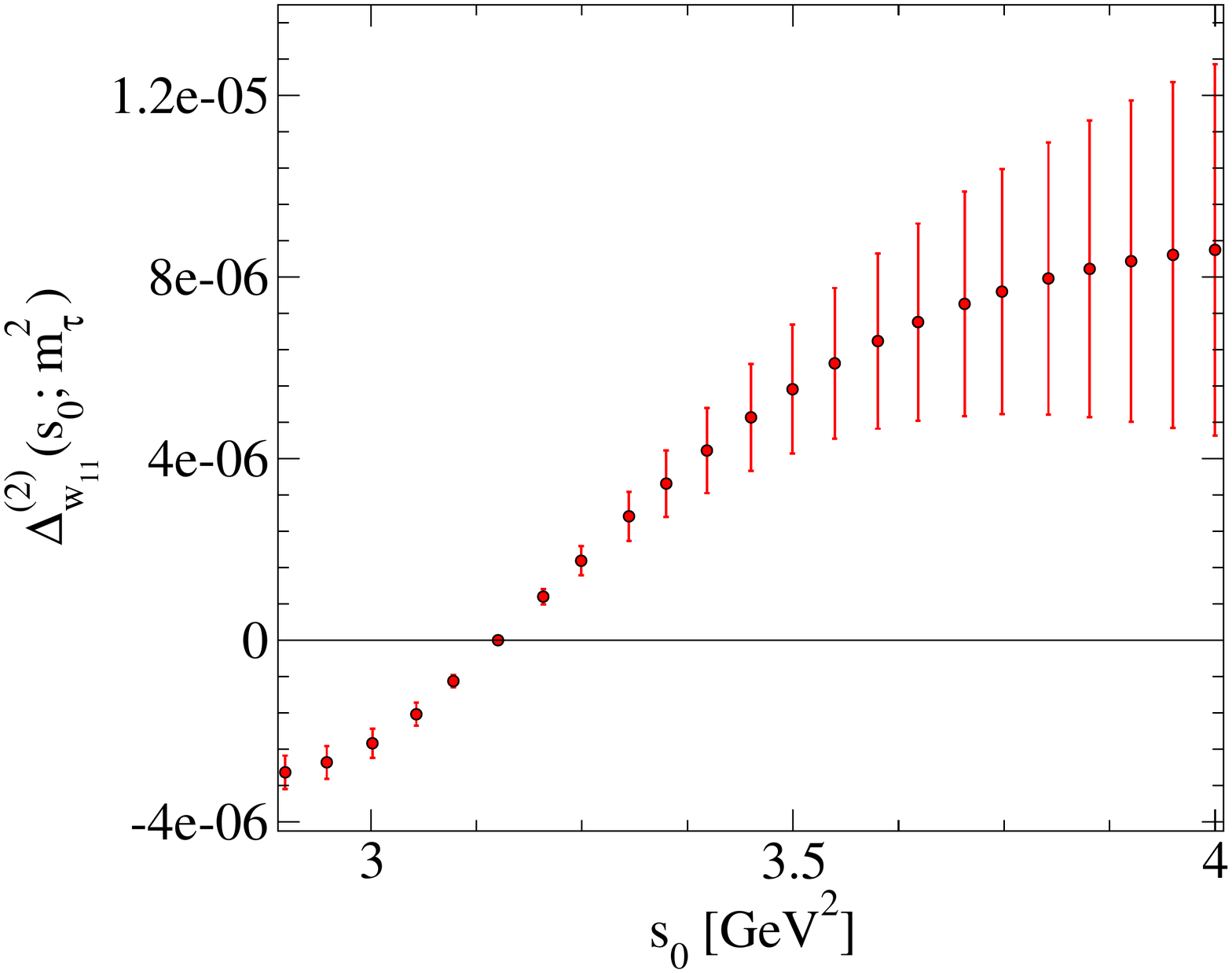}
\hspace{0.1cm}
\includegraphics*[width=7.4cm]{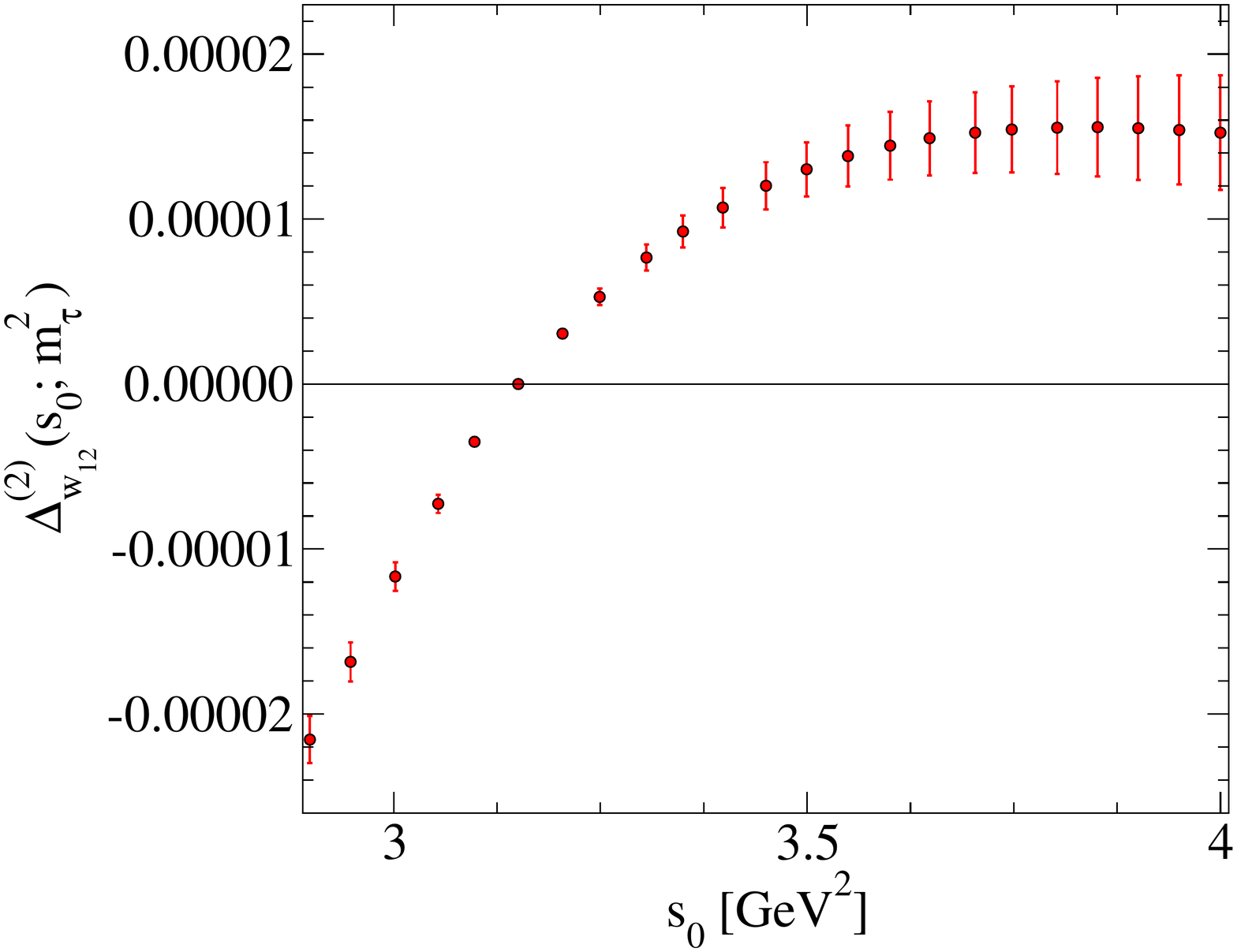}

\vspace{0.5cm}
\includegraphics*[width=7.4cm]{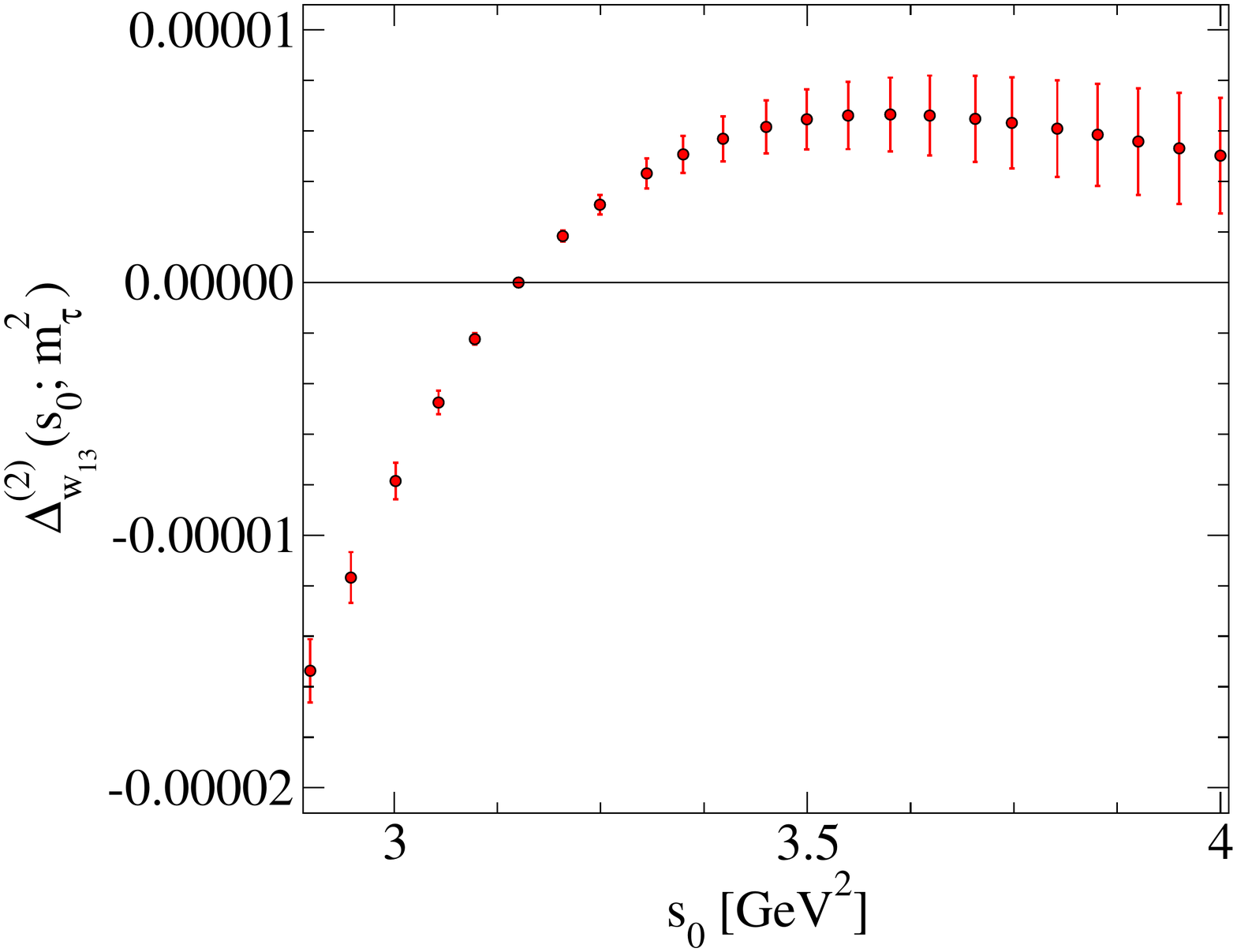}
\end{center}
\begin{quotation}
\floatcaption{classicaldiagdoublediff}%
{{\it The double differences, $\D^{(2)}_w(s_0;s_0^*)$, obtained from
diagonal fits with $k\ell$ spectral weights, as a function of $s_0$ with
$s_0^*=m_\t^2$.}}
\end{quotation}
\vspace*{-4ex}
\end{figure}

\begin{figure}[t]
\vspace*{4ex}
\begin{center}
\includegraphics*[width=7.4cm]{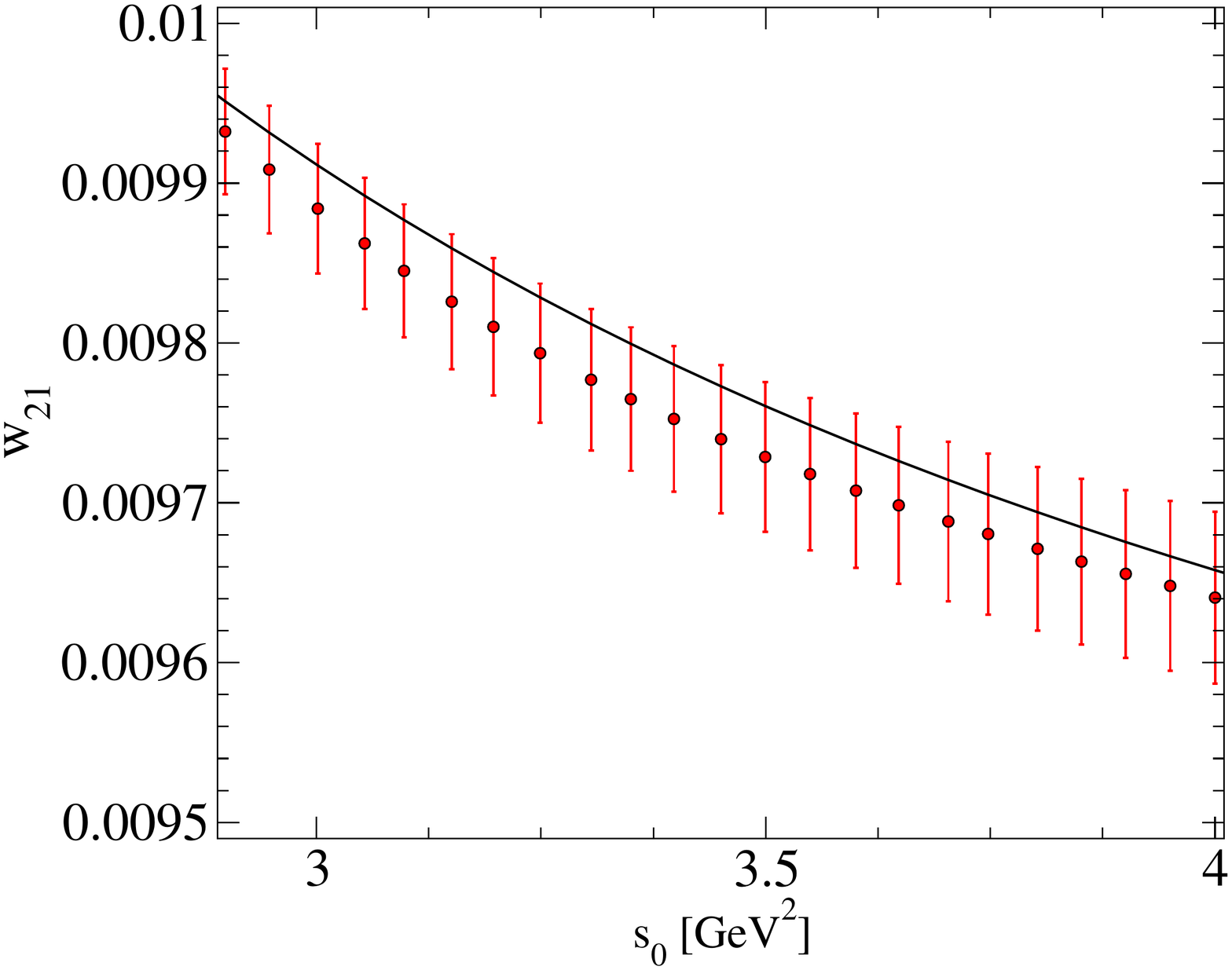}
\hspace{0.1cm}
\includegraphics*[width=7.4cm]{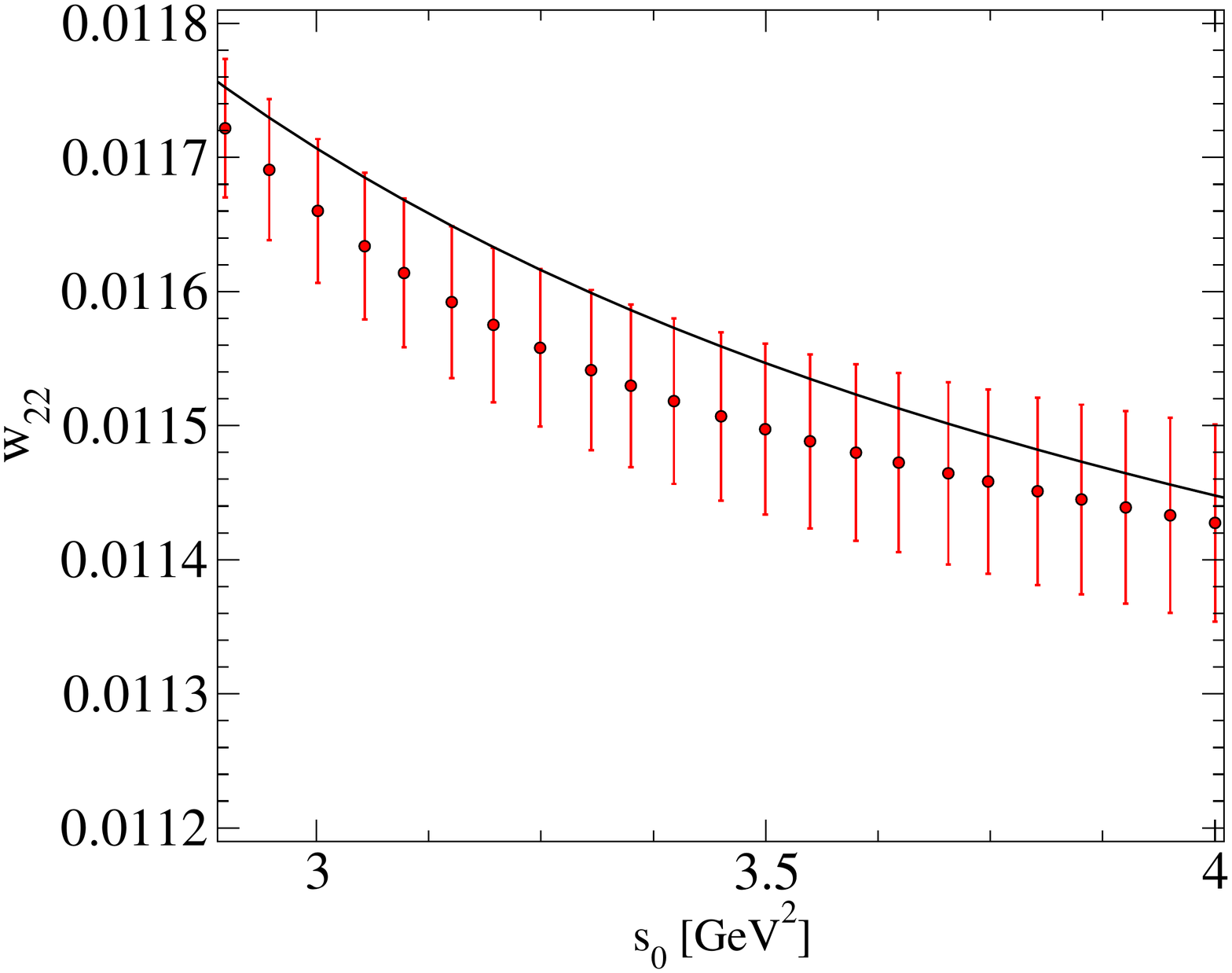}
\vspace{0.5cm}
\includegraphics*[width=7.4cm]{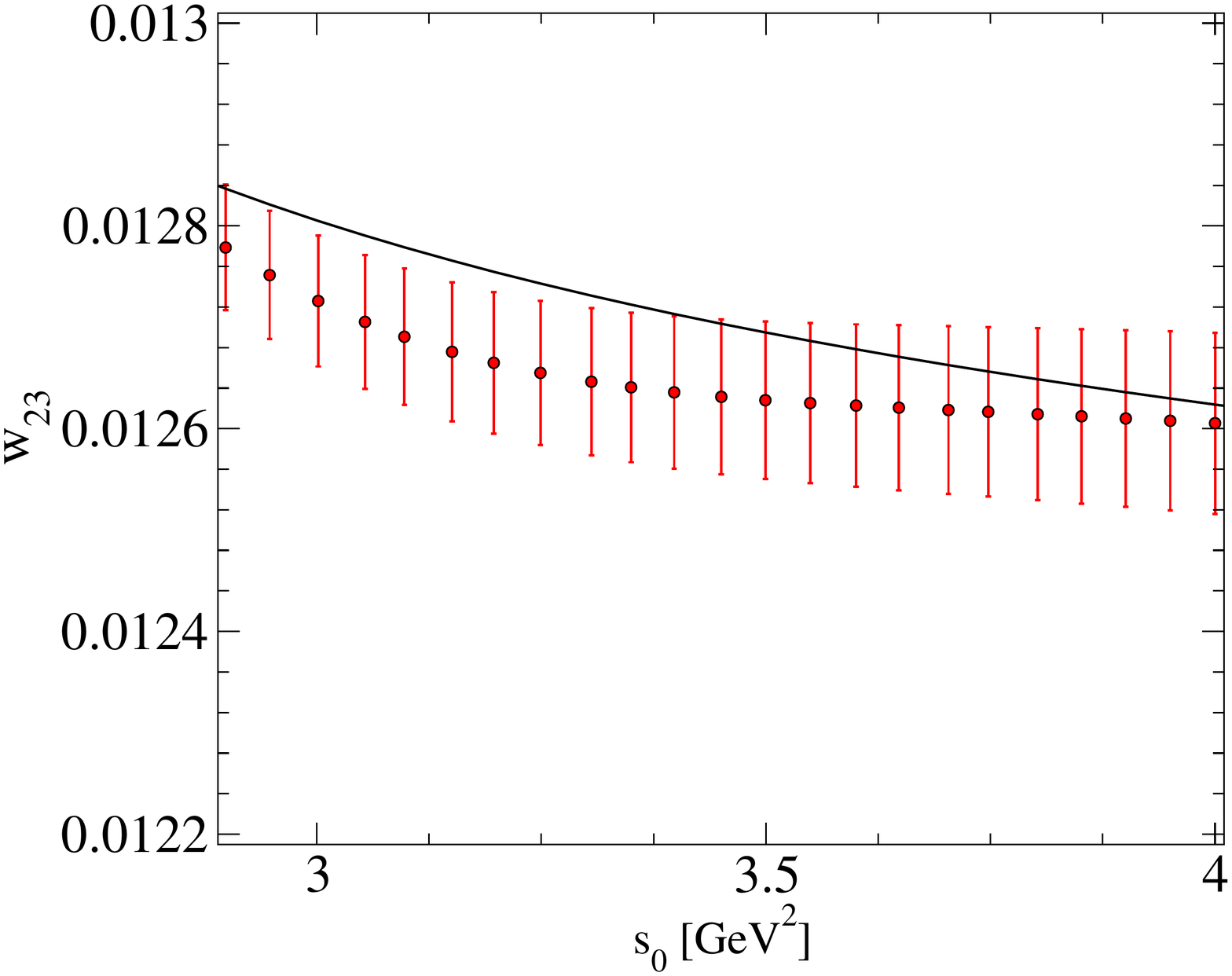}
\hspace{0.1cm}
\includegraphics*[width=7.4cm]{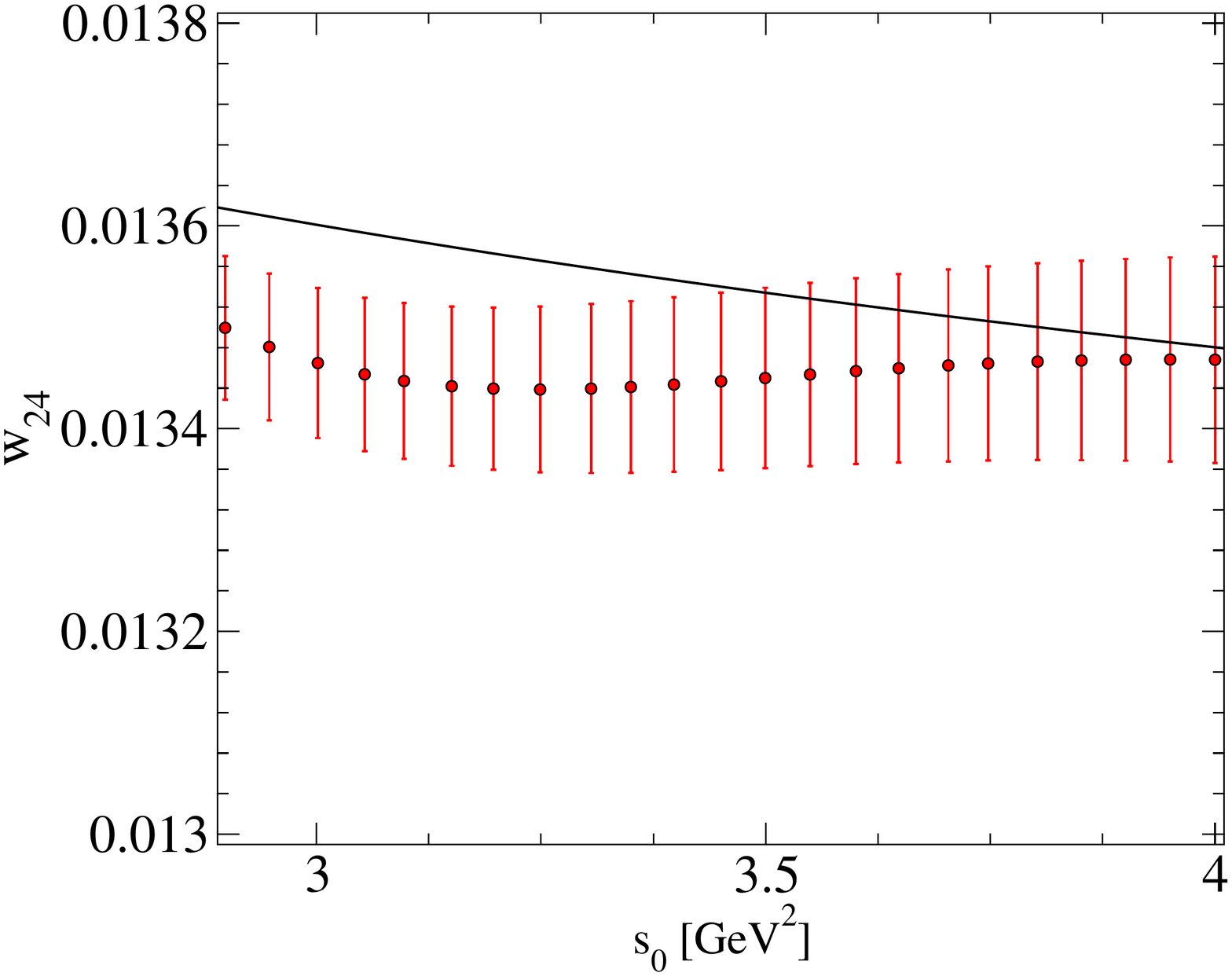}
\vspace{0.5cm}
\includegraphics*[width=7.4cm]{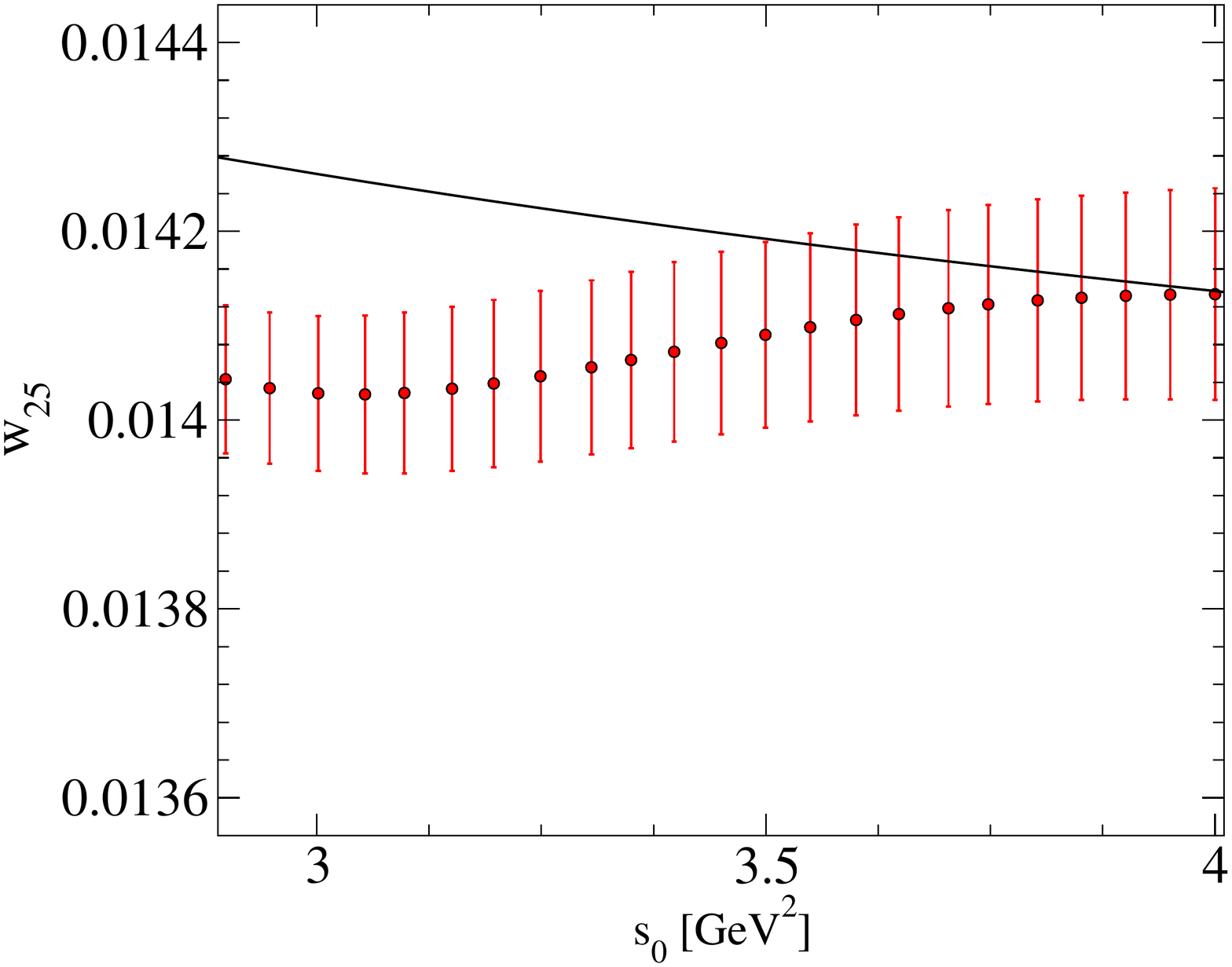}
\end{center}
\begin{quotation}
\floatcaption{optimalcorr}%
{{\it Comparison of $I_w^{\rm exp}(s_0)$ with $I_w^{\rm th}(s_0)$ with parameter
values obtained from correlated fits with optimal weights, as a function of
$s_0$ with $s_0^*=3.6$~{\rm GeV}$^2$.}}
\end{quotation}
\vspace*{-4ex}
\end{figure}

\begin{figure}[t]
\vspace*{4ex}
\begin{center}
\includegraphics*[width=7.4cm]{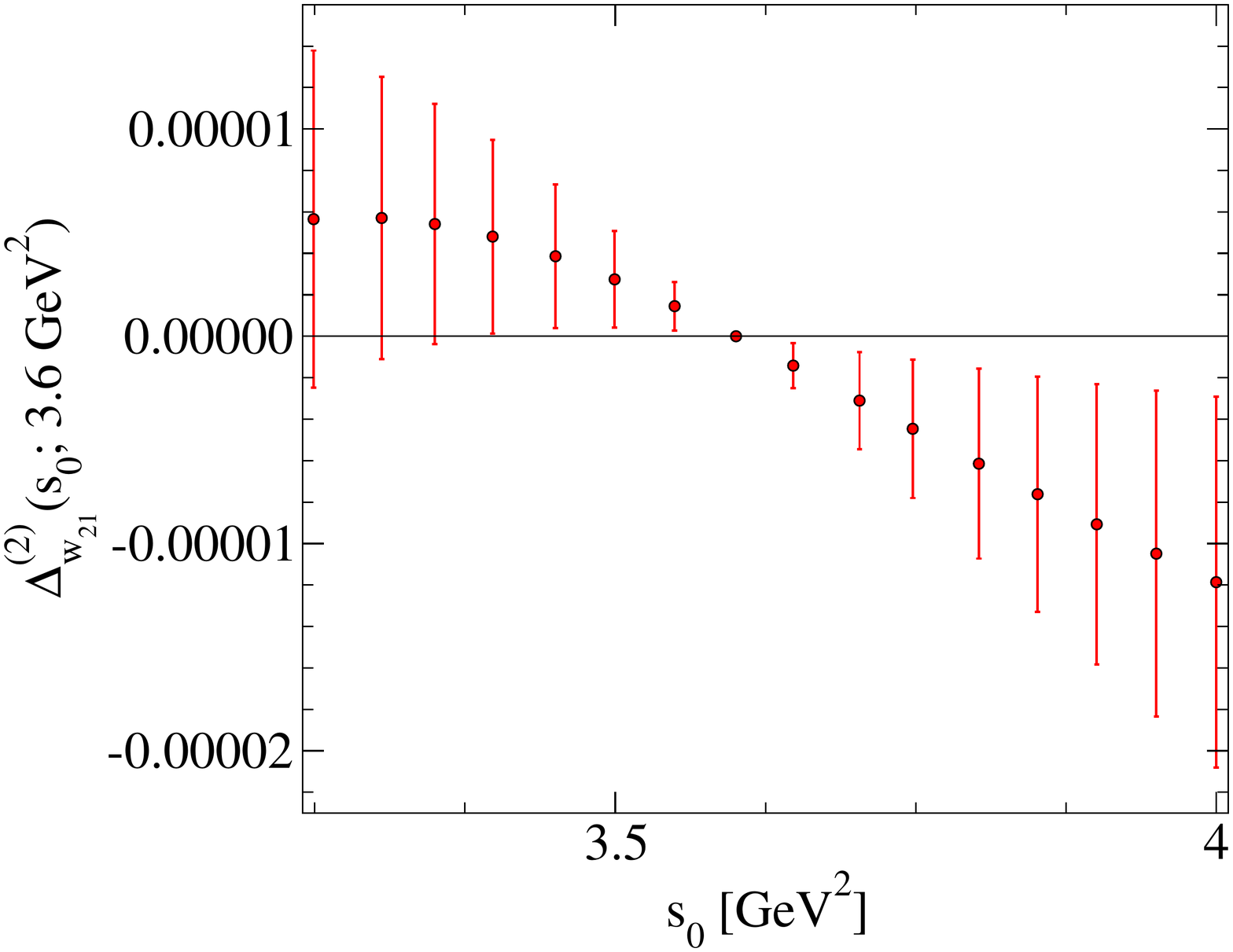}
\hspace{0.1cm}
\includegraphics*[width=7.4cm]{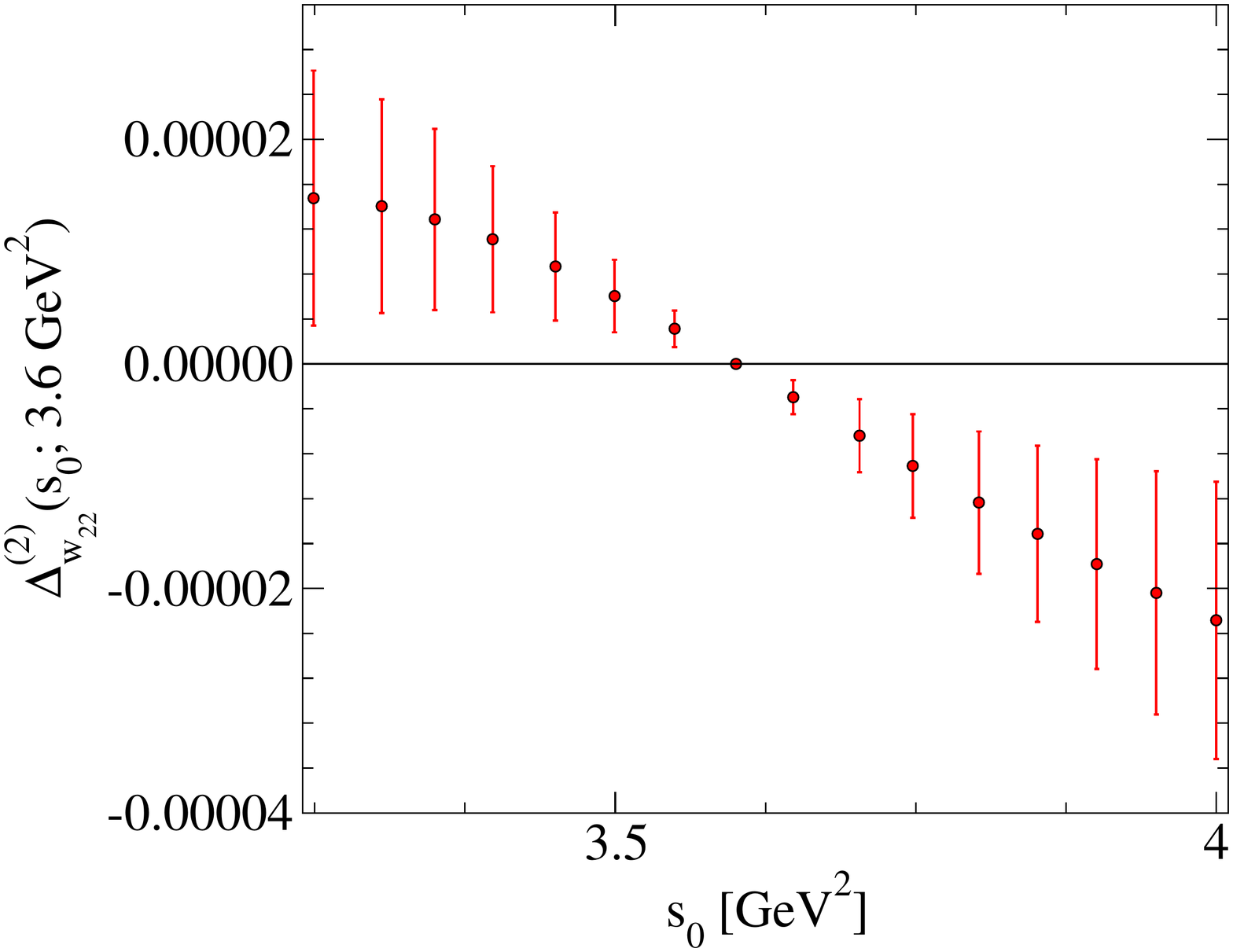}

\vspace{0.5cm}
\includegraphics*[width=7.4cm]{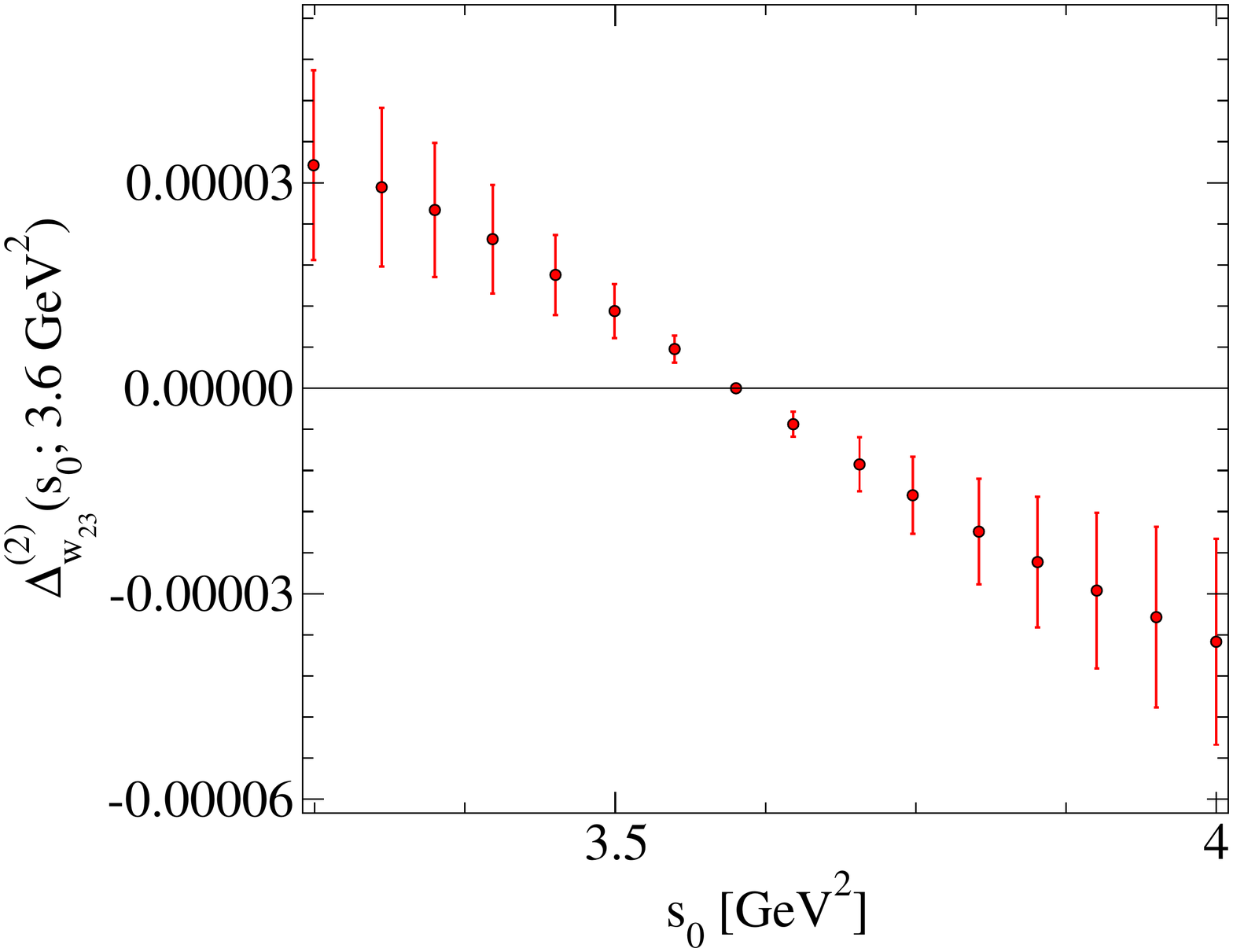}
\hspace{0.1cm}
\includegraphics*[width=7.4cm]{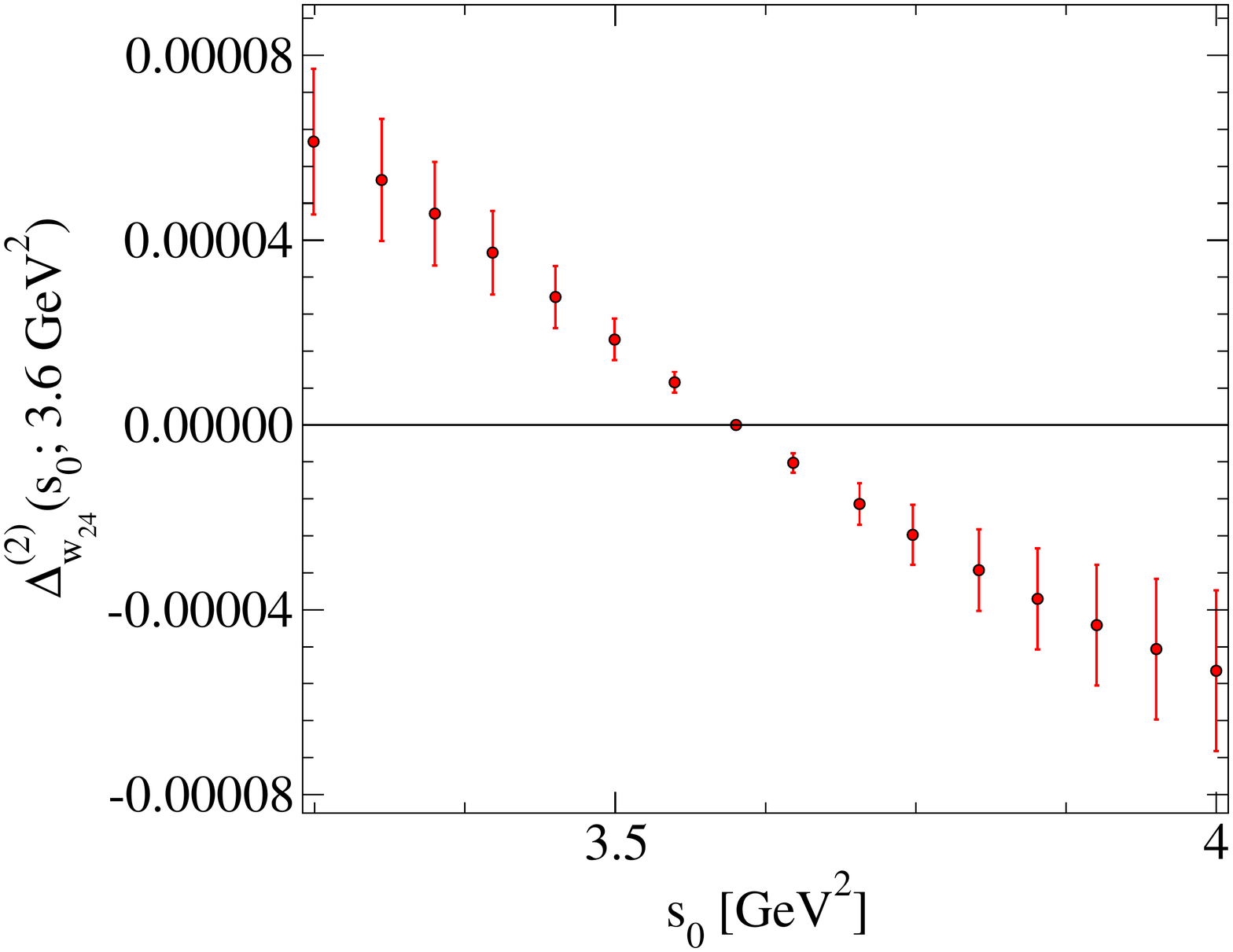}

\vspace{0.5cm}
\includegraphics*[width=7.4cm]{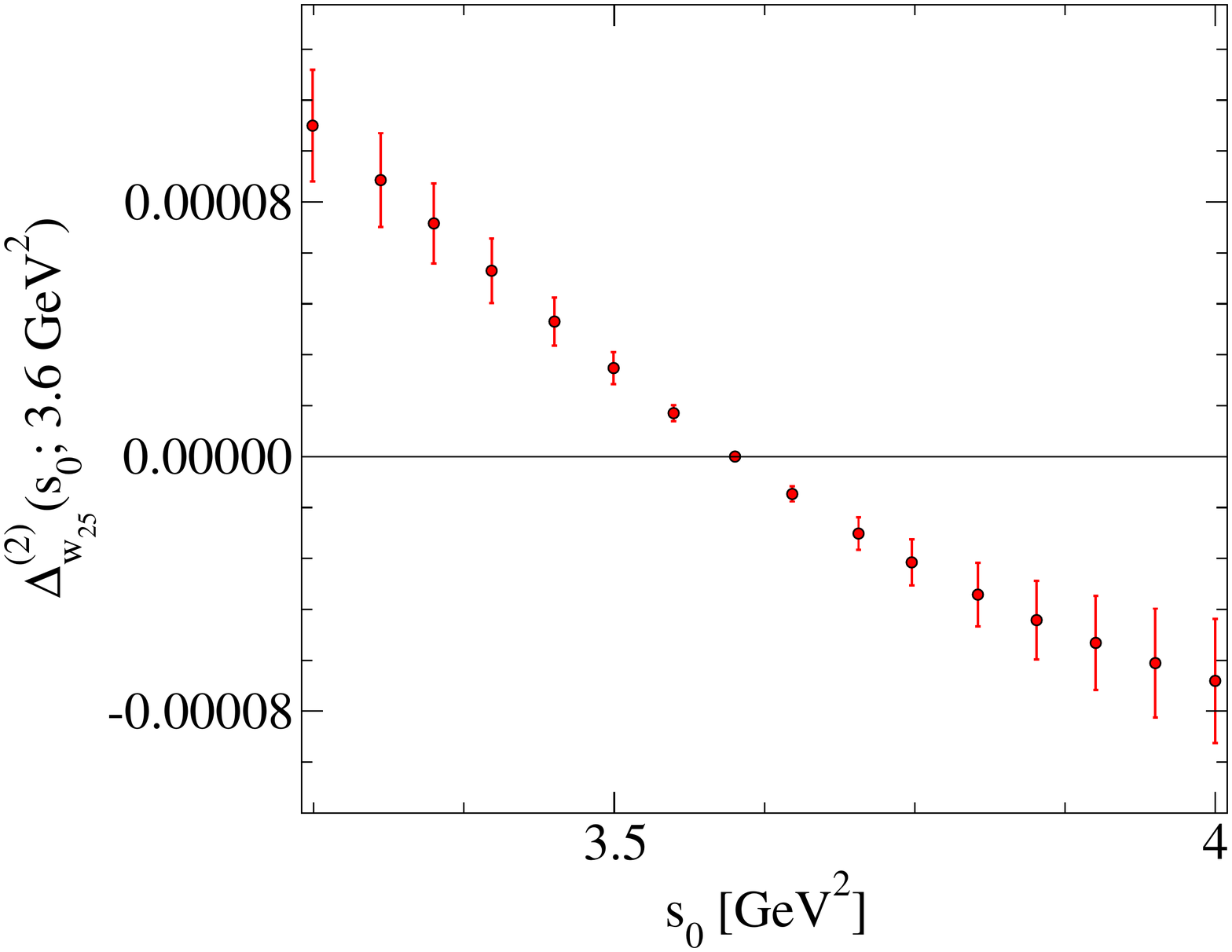}
\end{center}
\begin{quotation}
\floatcaption{optimalcorrdoublediff}%
{{\it The double differences, $\D^{(2)}_w(s_0;s_0^*)$, obtained from
correlated fits with optimal weights, as a function of $s_0$ with
$s_0^*=3.6$~{\rm GeV}$^2$.}}
\end{quotation}
\vspace*{-4ex}
\end{figure}

\begin{figure}[t]
\vspace*{4ex}
\begin{center}
\includegraphics*[width=7.4cm]{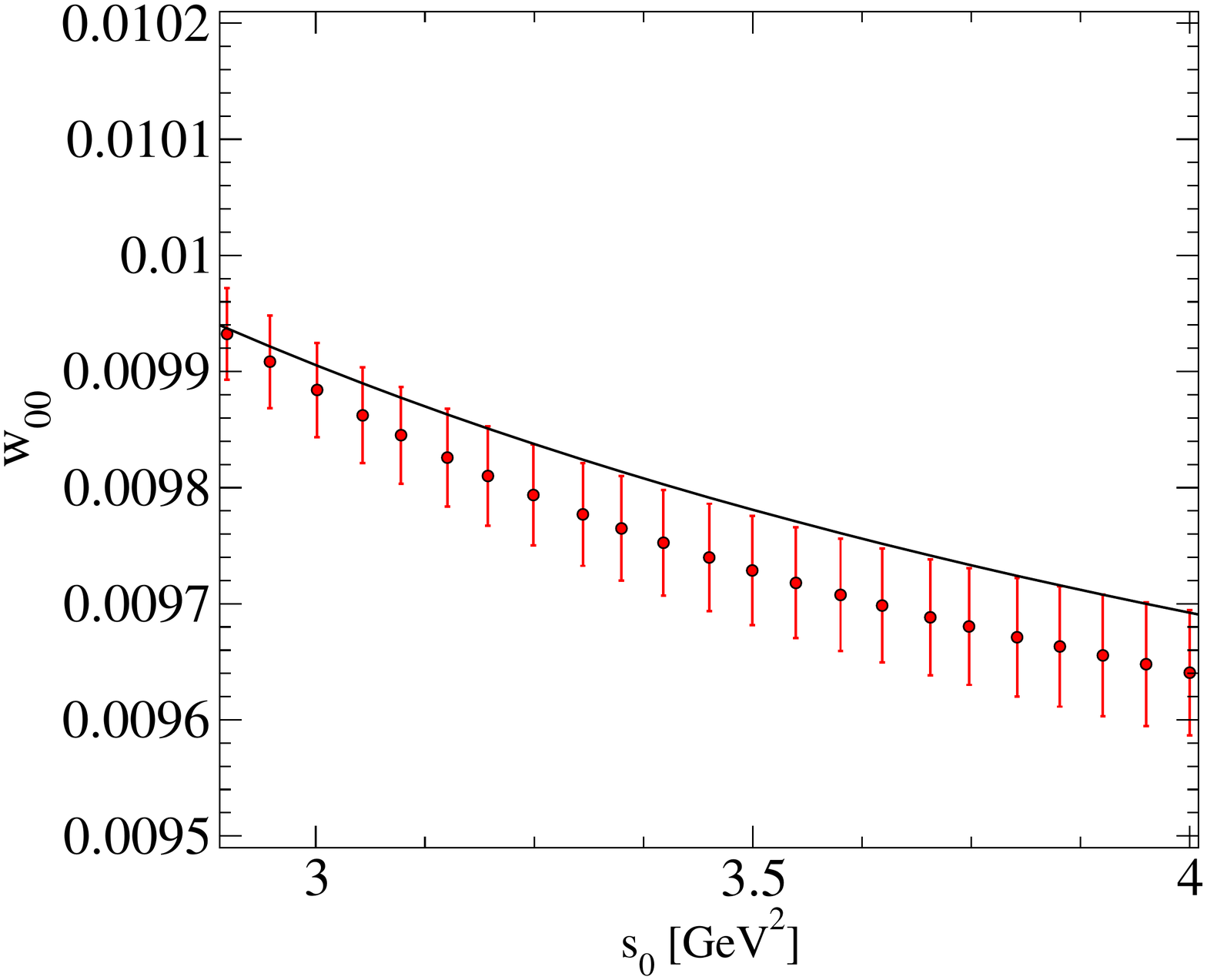}
\hspace{0.1cm}
\includegraphics*[width=7.4cm]{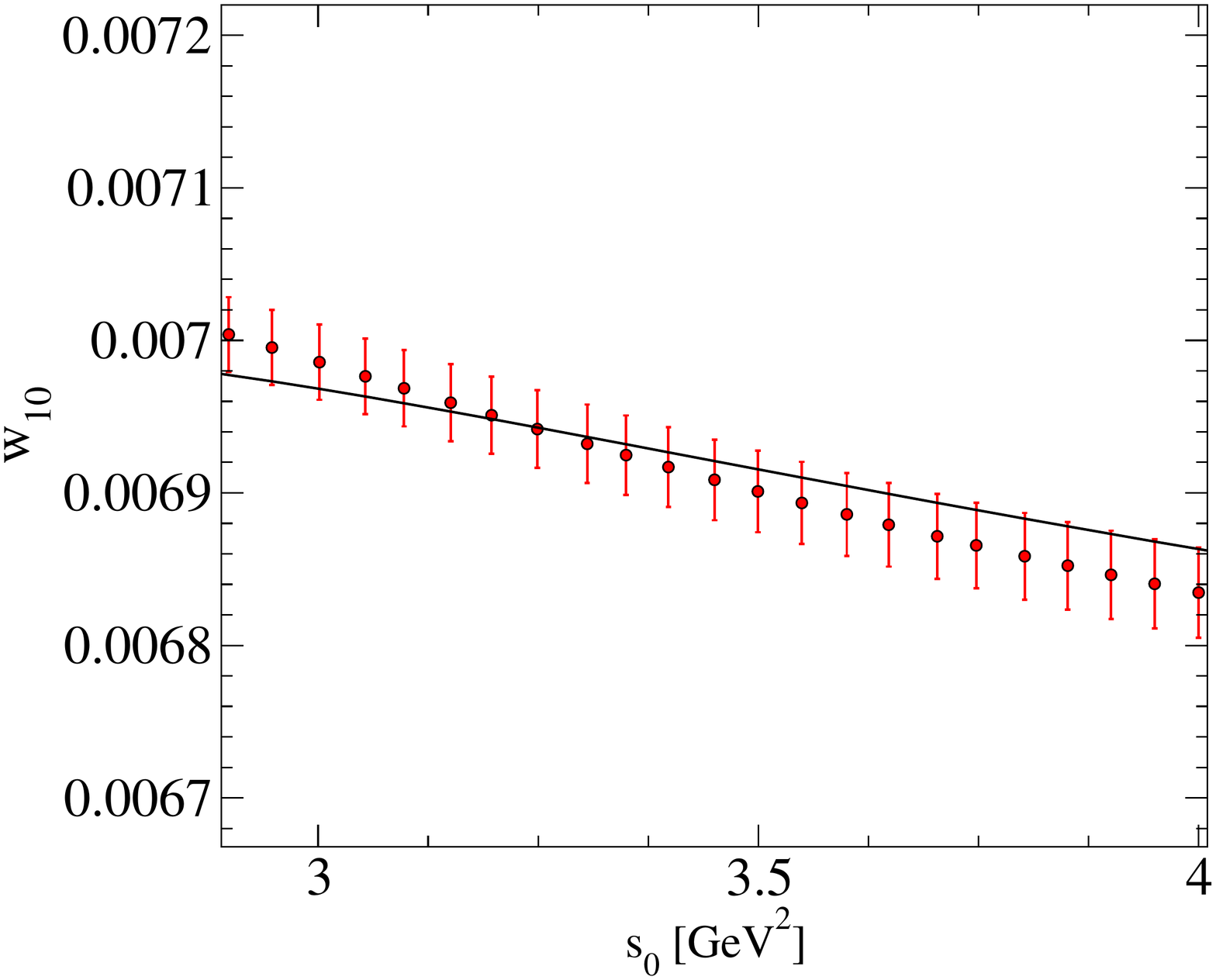}

\vspace{0.5cm}
\includegraphics*[width=7.4cm]{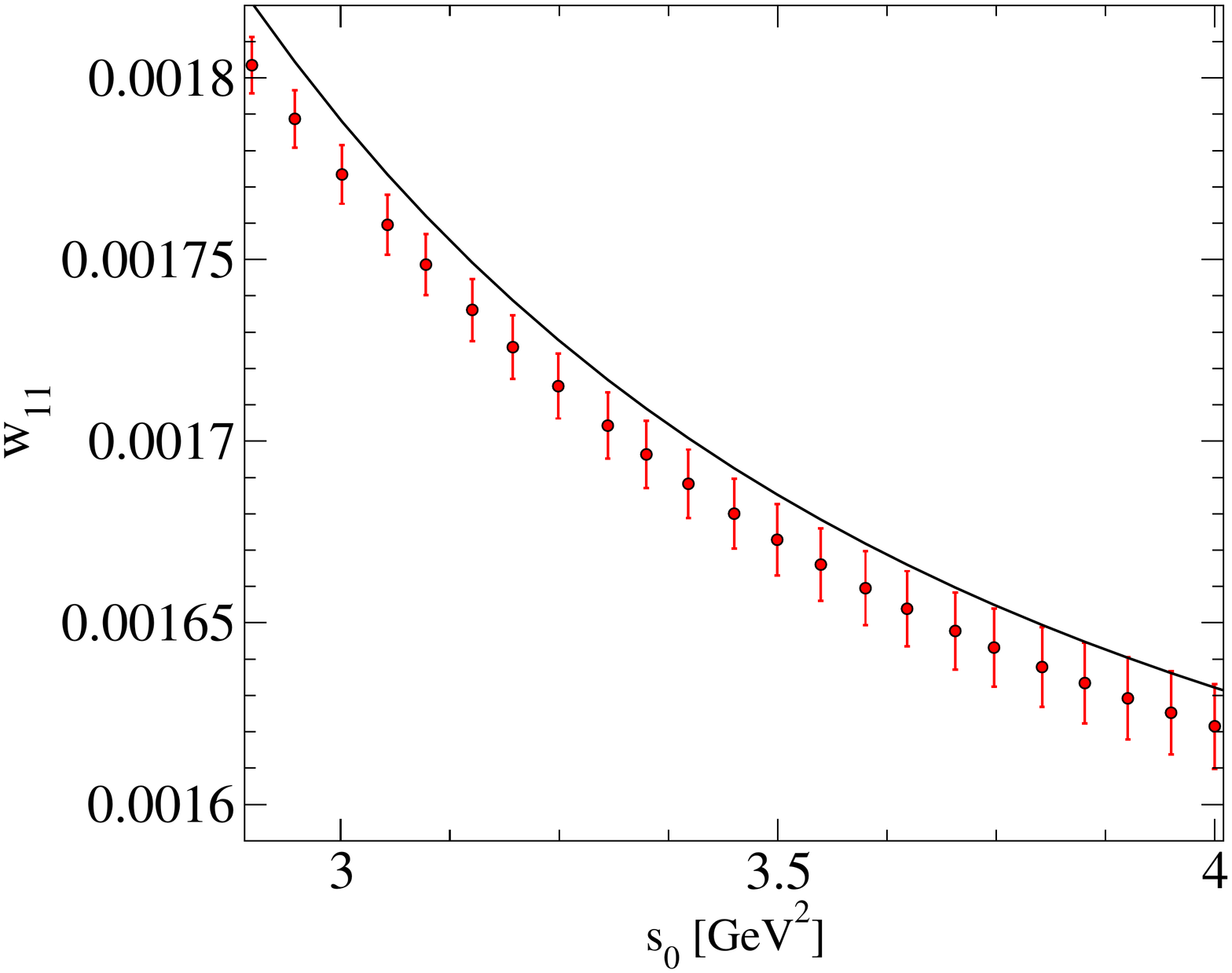}
\hspace{0.1cm}
\includegraphics*[width=7.4cm]{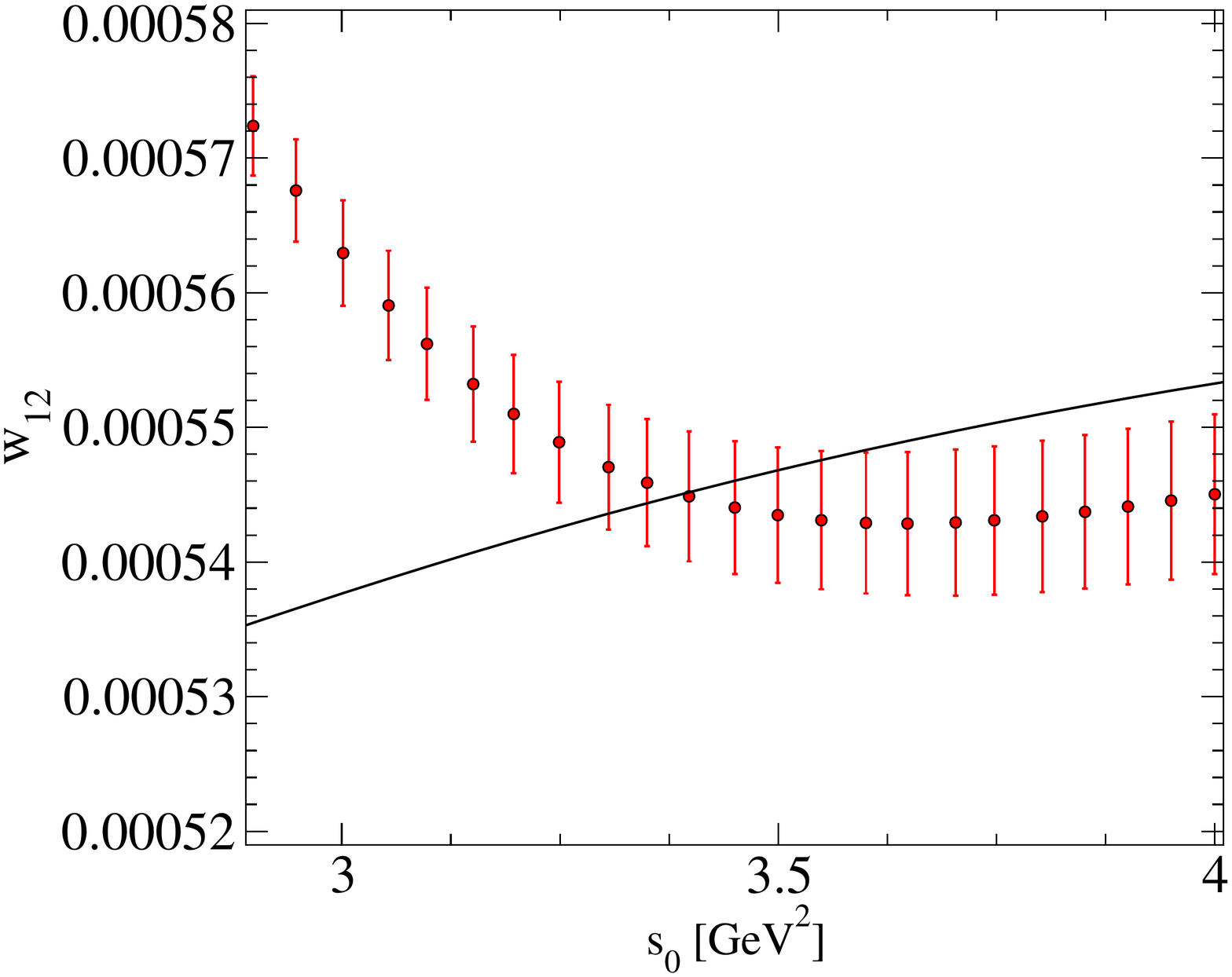}

\vspace{0.5cm}
\includegraphics*[width=7.4cm]{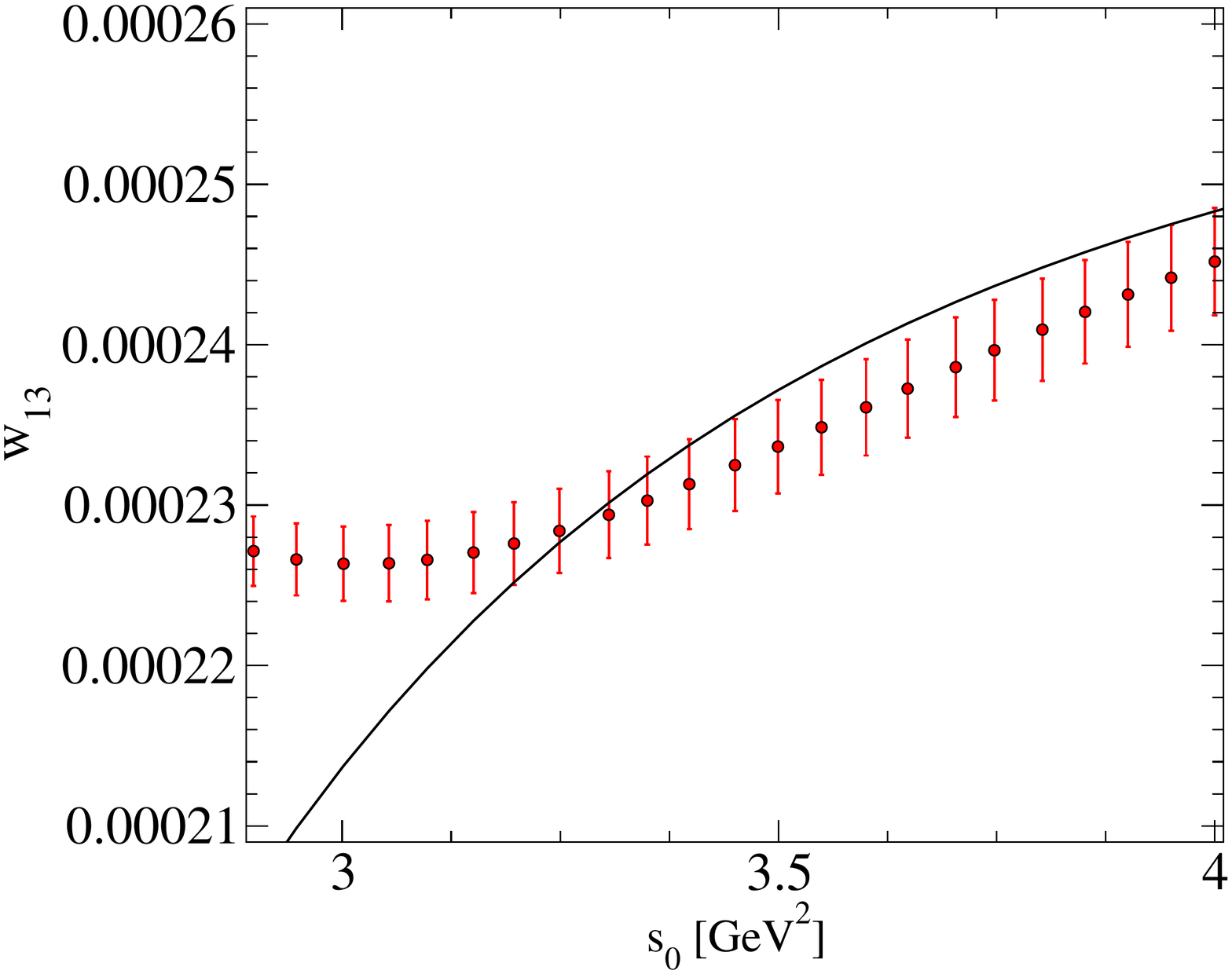}
\end{center}
\begin{quotation}
\floatcaption{classicalcorr}%
{{\it Comparison of $I_w^{\rm exp}(s_0)$ with $I_w^{\rm th}(s_0)$ with parameter
values obtained from correlated fits with $k\ell$ spectral weights, as a
function of $s_0$ with $s_0^*=3.7$~{\rm GeV}$^2$.}}
\end{quotation}
\vspace*{-4ex}
\end{figure}

\begin{figure}[t]
\vspace*{4ex}
\begin{center}
\includegraphics*[width=7.4cm]{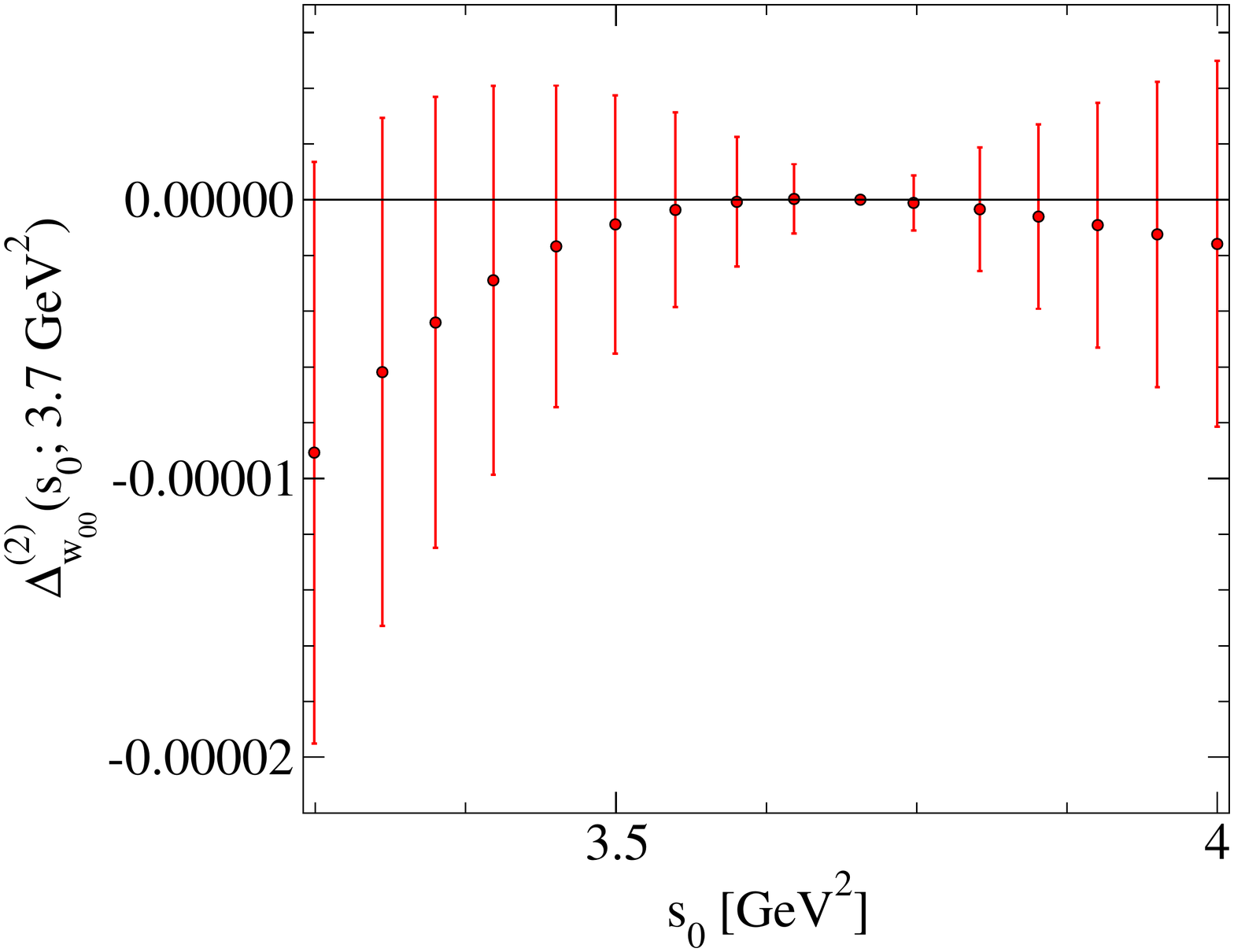}
\hspace{0.1cm}
\includegraphics*[width=7.4cm]{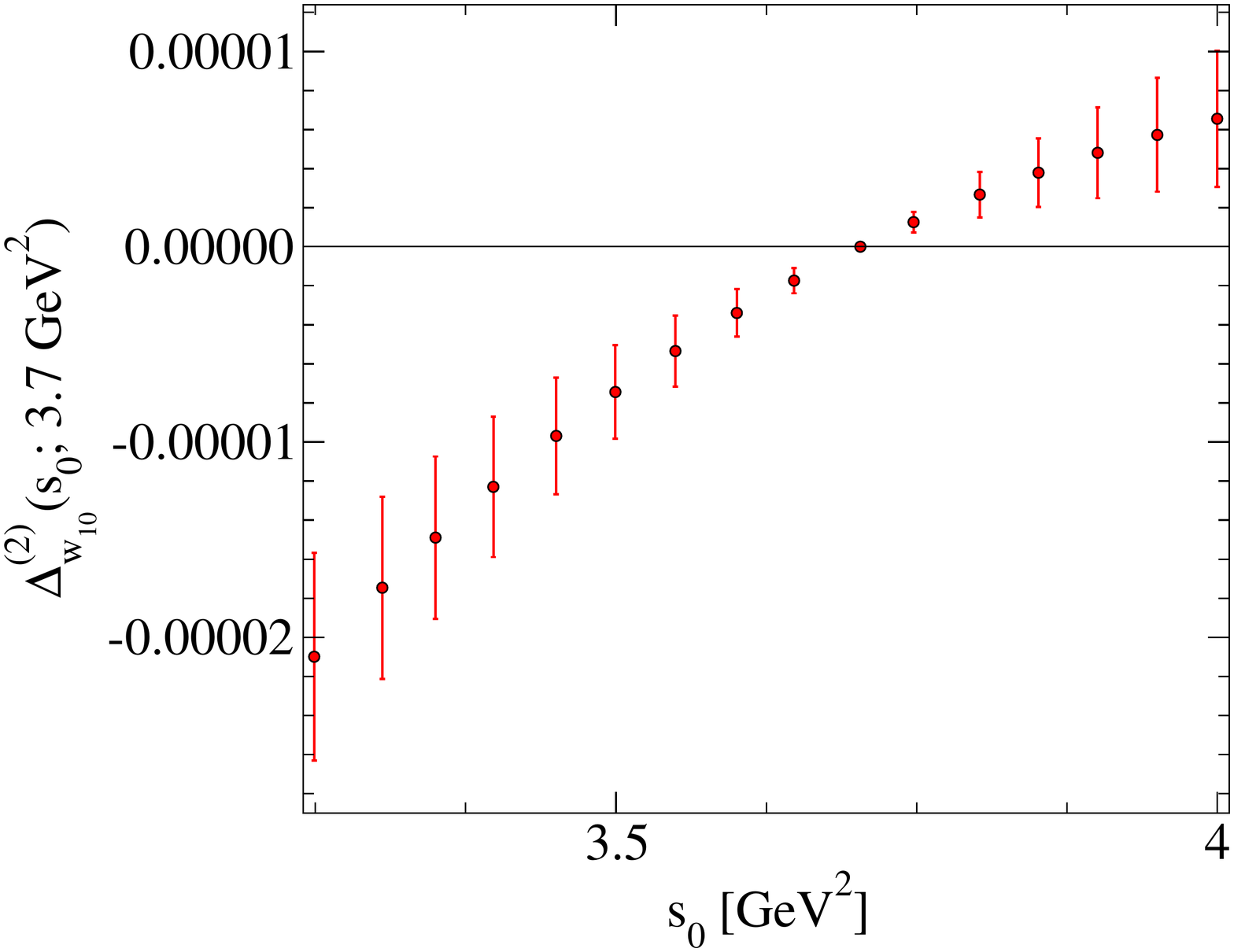}
\vspace{0.5cm}
\includegraphics*[width=7.4cm]{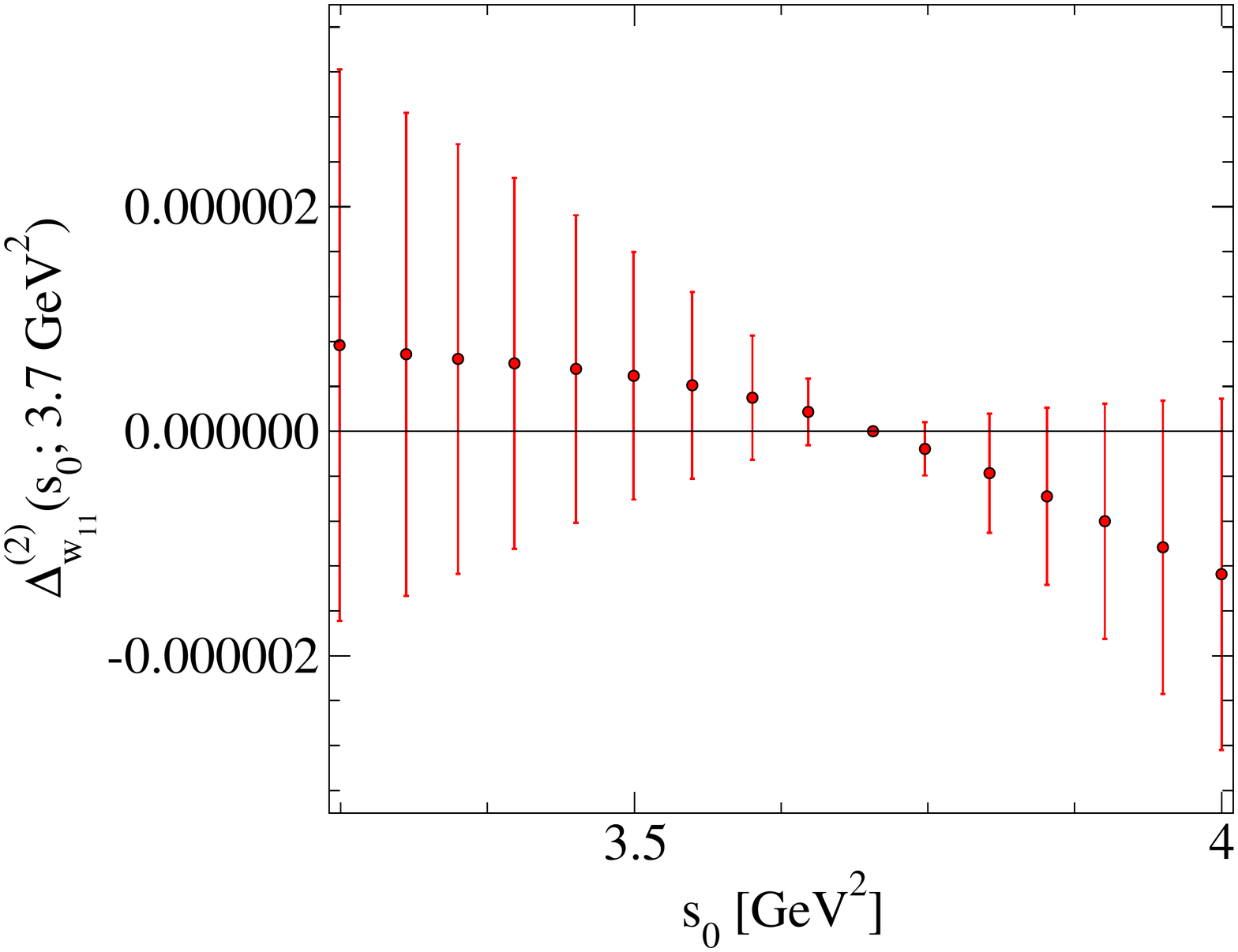}
\hspace{0.1cm}
\includegraphics*[width=7.4cm]{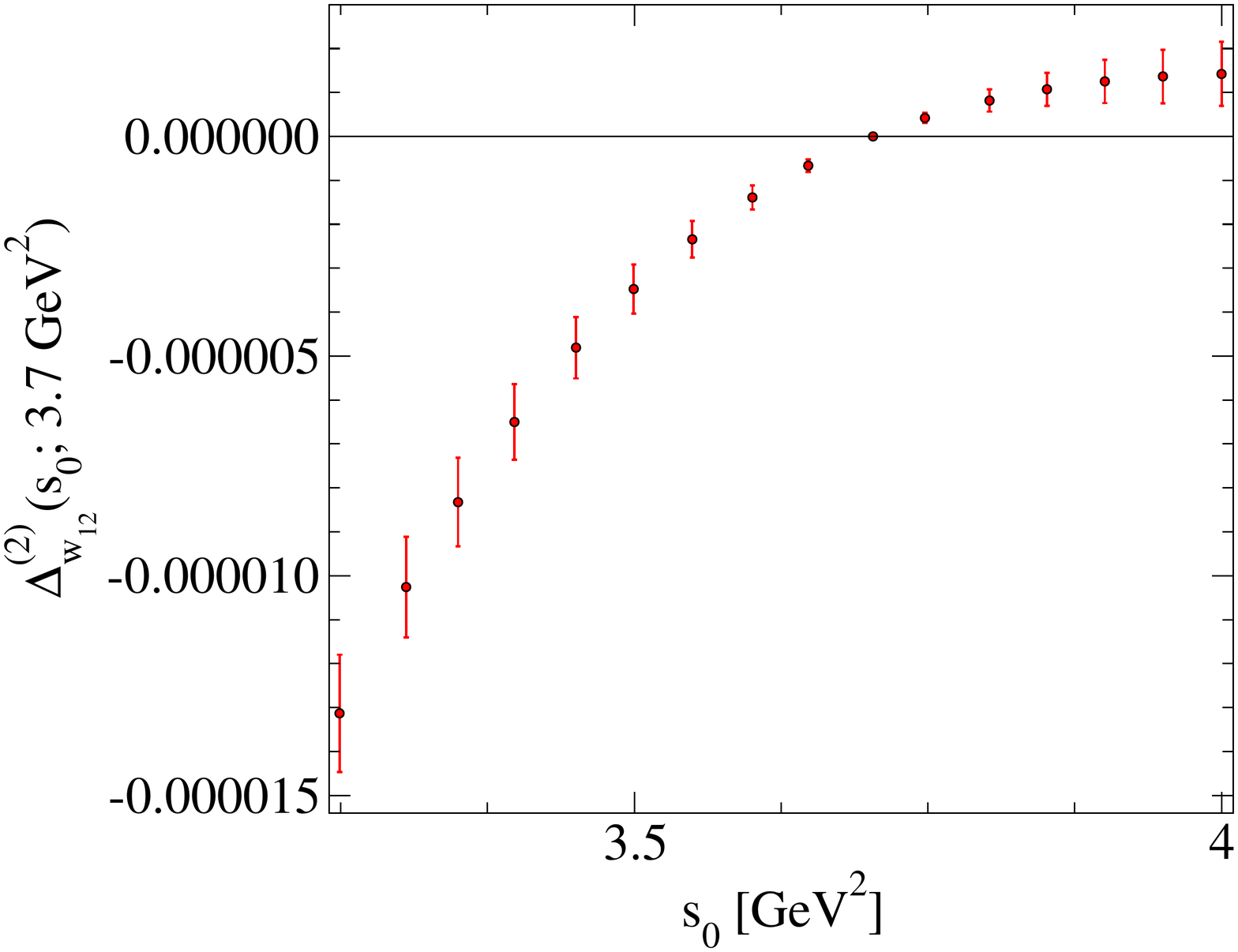}

\vspace{0.5cm}
\includegraphics*[width=7.4cm]{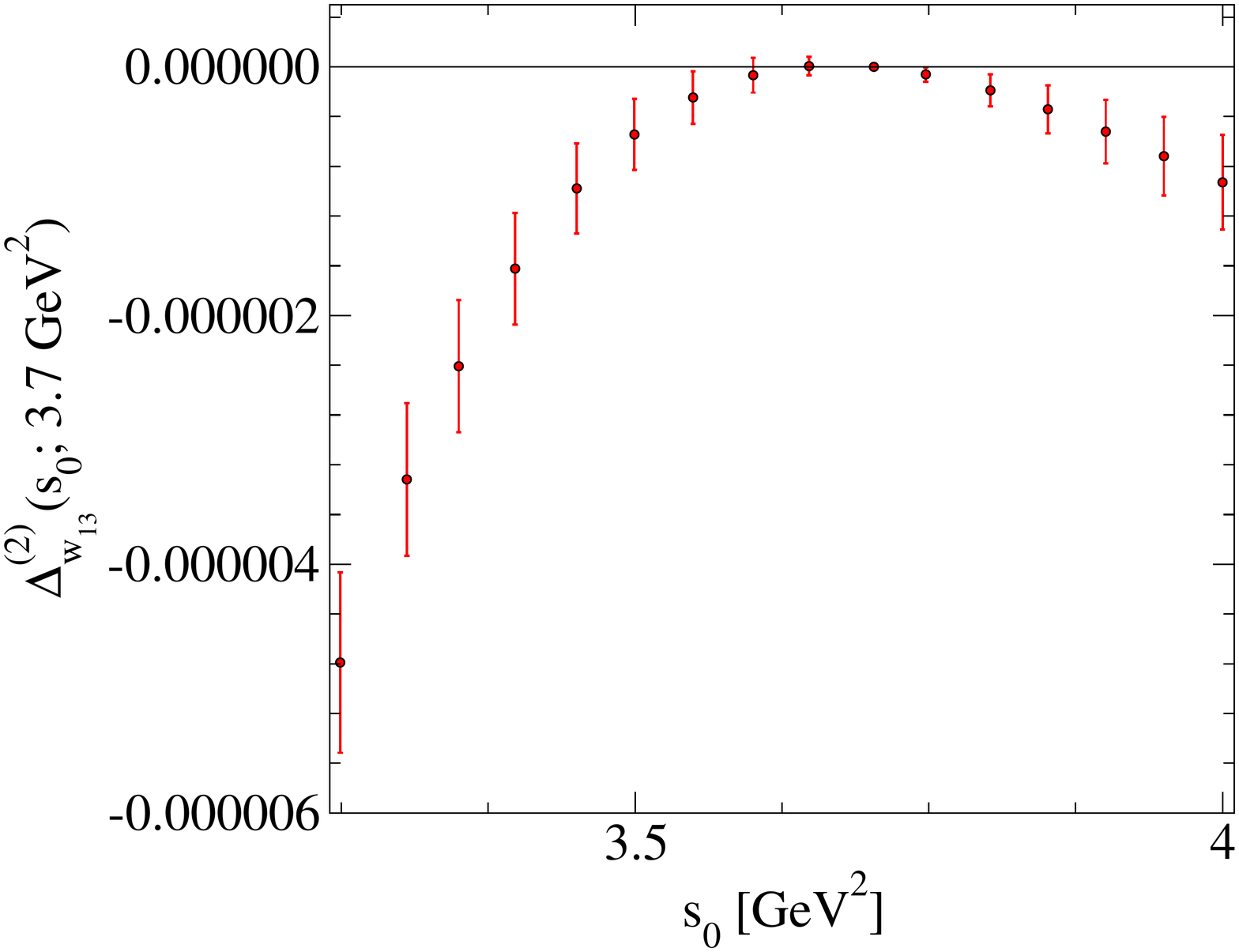}
\end{center}
\begin{quotation}
\floatcaption{classicalcorrdoublediff}%
{{\it The double differences, $\D^{(2)}_w(s_0;s_0^*)$, obtained from
correlated fits with $k\ell$ spectral weights, as a function of $s_0$ with
$s_0^*=3.7$~{\rm GeV}$^2$.}}
\end{quotation}
\vspace*{-4ex}
\end{figure}

\end{document}